\documentclass[a4wide,11pt]{article}

\pdfoutput=1 
\usepackage{amsfonts,amssymb,amsmath,amsthm,mathrsfs,mathtools}
\usepackage{graphicx,graphics}
\usepackage[T1]{fontenc}
\usepackage{cite,appendix}
\usepackage{fancyhdr,layout,lastpage}
\usepackage{longtable,upquote}
\usepackage{subcaption}
\usepackage{pstricks}
\usepackage{empheq}
\usepackage{lipsum}

\usepackage{jheppub}

\usepackage{color}
\usepackage{xcolor}
\definecolor{urlblue}{rgb}{0.2,0.4,0.7}
\definecolor{citegreen}{rgb}{0,0.6,0.2}
\definecolor{linkred}{rgb}{0.9,0.2,0.1}
\usepackage{hyperref}
\hypersetup{colorlinks=true, citecolor=citegreen, linkcolor=blue, urlcolor=blue}

\def\MSbar{\overline{\mathrm{MS}}}
\def\als{\alpha_s}
\def\as{a_s}

\newcommand{\Oals}[1]{\mathcal{O}(\alpha_s^{#1})}

\def\mtms{\overline{m}_t}

\def\mtkin{m_{t,\mathrm{kin}}}

\def\mbms{\overline{m}_b}
\def\mbos{m_{b,\mathrm{os}}}
\def\mbkin{m_{b,\mathrm{kin}}}
\def\mbOs{m_{b,\mathrm{1s}}}
\def\mbsm{m_{b,\sigma}}

\def\mcms{\overline{m}_c}
\def\mcos{m_{c,\mathrm{os}}}
\def\mckin{m_{c,\mathrm{kin}}}
\def\mcOs{m_{c,\mathrm{1s}}}
\def\mcsm{m_{c,\sigma}}

\def\Gamt{\mathrm{\Gamma}_t}
\def\HFs{f_{\mathrm{L,R,0}}}

\def\HTensor{\mathcal{W}_{Qq}^{\,\mu \nu}}

\def\BXulv{B \rightarrow X_u \ell \bar{\nu}_{\ell}}
\def\bulv{b \rightarrow u \ell \bar{\nu}_{\ell}}
\def\GamBulv{\Gamma^{\mathrm{sl}}_{b\rightarrow u}}
\def\Bpilv{B \rightarrow \pi \ell \bar{\nu}_{\ell}}

\def\cdlv{c \rightarrow d\,\ell \bar{\nu}_{\ell}}

\def\cqlv{c \rightarrow q\,\ell \bar{\nu}_{\ell}}
\def\Dqlv{D \rightarrow X\,\ell \bar{\nu}_{\ell}}

\def\Gamsl{\Gamma^{\mathrm{sl}}}
\def\GamTsl{\Gamma^{\mathrm{sl}}_{t}}

\def\GamBulvMS{\bar{\Gamma}^{\mathrm{sl}}_{b\rightarrow u}}
\def\GamBulvK{\tilde{\Gamma}^{\mathrm{sl}}_{b\rightarrow u}}

\def\BXclv{B \rightarrow X_c \ell \bar{\nu}_{\ell}}

\title{
Heavy-to-light Structure Functions at $\mathcal{O}(\alpha_s^3)$ in QCD %%%
}

\author[a]{Long Chen,}
\emailAdd{longchen@sdu.edu.cn}

\author[b]{Xiang Chen,}
\emailAdd{xiang.chen@physik.uzh.ch}

\author[c]{Xin Guan,}
\emailAdd{guanxin@slac.stanford.edu}

\author[d,e]{Yan-Qing Ma}
\emailAdd{yqma@pku.edu.cn}

\affiliation[a]{School of Physics, Shandong University, Jinan, Shandong 250100, China}
\affiliation[b]{Physik-Institut, Universit\"at Z\"urich, Winterthurerstrasse 190, CH-8057 Z\"urich, Switzerland}
\affiliation[c]{SLAC National Accelerator Laboratory, Stanford University, Stanford, California 94039, USA}
\affiliation[d]{School of Physics, Peking University, Beijing 100871, China}
\affiliation[e]{Center for High Energy Physics, Peking University, Beijing 100871, China}

\abstract{
We present the first complete $\mathcal{O}(\alpha_s^2)$ and $\mathcal{O}(\alpha_s^3)$ perturbative QCD corrections to all five heavy-to-light structure functions underlying the triple-differential semi-leptonic decay rates of heavy quarks. This is achieved via a hybrid computational strategy that combines an efficient linear interpolation (with a suitable function basis) based on stratified Gauss-Kronrod points in the leptonic-mass $q^2$ with the differential equations in the other variable, further armed with reduced numerical $\varepsilon$-dependence. Among the selected applications, we highlight the state-of-the-art prediction $\Gamma(B \rightarrow X_u \ell \bar{\nu}_{\ell}) = \frac{|V_{ub}|^2}{|3.82\times 10^{-3}|^2}\,\big( 6.53 \,\pm 0.12 \, \pm 0.13\, \pm 0.03\, \big) \times 10^{-16}\,\text{GeV}\,$ derived in the kinetic-mass scheme. We report several notable observations regarding the convergence of the first three orders of QCD corrections to the $q^2$-spectrum and to inclusive moments of the lepton-energy spectrum in semi-leptonic weak decays of $b$- and $c$-quark in different quark-mass schemes; they are important both for improving the inclusive determinations of the relevant CKM elements, non-perturbative dynamical parameters, and for gaining new insights into the potential impact of high-order QCD corrections. Lastly we discuss a novel interesting point encountered in the consistent perturbative reformulation of the differential $q^2$-spectrum from the pole-mass to other mass schemes: certain boundary-effect terms are identified that are non-vanishing for $b \rightarrow u \ell \bar{\nu}_{\ell}$ firstly at $\mathcal{O}(\alpha_s^3)$; their incorporation is essential to preserve the integrity of the integrated moments of the perturbatively re-expanded $q^2$-spectrum but necessitates histogramming from $\mathcal{O}(\alpha_s^3)$ onward even within pure perturbation theory.
}

\begin{document}
\preprint{CPTNP-2026-007, SLAC-PUB-260209-2, ZU-TH 07/26}

\allowdisplaybreaks[4]
\unitlength1cm
\maketitle
\flushbottom

\section{Introduction}\label{sec:intro}

The semi-leptonic decays of heavy quarks or heavy-flavored hadrons constitute a unique natural laboratory for probing fundamental aspects of the Standard Model, including the Cabibbo–Kobayashi–Maskawa (CKM) mechanism, the dynamics of Quantum Chromodynamics (QCD), and providing access to the masses and decay widths of heavy quarks as well as the CKM matrix elements.
The current experimental status is driven by the ever-more precise data for $t$-quark studies at ATLAS and CMS~\cite{ATLAS:2008xda,CMS:2008xjf,ATLAS:2022hsp}, for $b$-quark physics from Belle II~\cite{Belle-II:2010dht,Belle-II:2018jsg} and LHCb~\cite{LHCb:2018roe}, and for $c$-quark physics from BES III~\cite{BESIII:2009fln,BESIII:2020nme} --- advancing into percent-level~\cite{HFLAV:2022esi,HeavyFlavorAveragingGroupHFLAV:2024ctg} --- with even higher precision anticipated from future colliders~\cite{FCC:2018evy,CLICdp:2018esa,Belle-II:2022cgf,Ai:2024nmn}. 
The (perturbative) QCD corrections are among the essential theoretical ingredients both for connecting the measured hadronic observables to theory parameters defined in fundamental or effective theories, and for reducing theoretical uncertainties to the levels commensurate with experimental precision. 
The ultimate goal remains to compress theoretical uncertainties to such a degree that any persistent discrepancy with experiments %beyond the experimental errors 
can be unambiguously ascribed as a sign of New Physics.
To this end, the continued refinement of perturbative QCD corrections to these processes is not merely a technical exercise; 
it is the essential process of sharpening our probes to test the Standard Model and seek for the beyond.
~\\

In the case of $t$-quark, the current precise measurement for the $t$-quark decay width $\Gamt$ comes from the CMS \cite{CMS:2014mxl} which gives $1.36\pm0.02\mathrm{(stat.)}^{+0.14}_{-0.11}\mathrm{(syst.)}\,$GeV.
The anticipated experimental uncertainties in the measurement of $\Gamt$ at the future hadron and lepton colliders can be reduced to about $20 \sim 26\,$MeV\cite{Horiguchi:2013wra,CLICdp:2018esa,Baskakov:2018huw,Li:2022iav}. 
To fully take advantage of the data, the theoretical error should be at least smaller than one-third of the experimental one,\footnote{In the same spirit as mentioned earlier in this Introduction, if the theoretical uncertainty is less than one third of the experimental one, then its contribution to the total uncertainty, combined by add-in-quadrature, will be almost an order-of-magnitude less than the latter one and hence becomes negligible in combination.} say less than $7\,$MeV.
Although QCD corrections to $t$-quark decay have been calculated perturbatively up to next-to-next-to leading order (N2LO) in the strong coupling $\as$, e.g.~in refs.\cite{Jezabek:1988iv,Czarnecki:1990kv,Li:1990qf,Czarnecki:1998qc,Chetyrkin:1999ju,Fischer:2001gp,Blokland:2004ye,Blokland:2005vq,Czarnecki:2010gb,Gao:2012ja,Brucherseifer:2013iv,Campbell:2020fhf,Meng:2022htg,Chen:2022wit}, the theoretical error at N2LO cannot meet this request.
In addition, there is also a more theoretically oriented motivation for an explicit determination of the next-to-next-to-next-to leading order (N3LO) QCD correction to the $t$-quark decay process: 
there are concerns over the convergence rate of the perturbative series related to the fact that both the pole mass $m_t$ and the decay width $\Gamt$ are sensitive to the infrared-renormalon issue (See~e.g.~refs.\cite{Bigi:1994em,Beneke:1994sw,Beneke:1998ui,Beneke:1994qe,Smith:1996xz,Beneke:1998ui,Beneke:2016cbu,Hoang:2017btd,FerrarioRavasio:2018ubr}).
It is thus desirable to explicitly determine the \textit{complete} QCD corrections at N3LO in $\als$ to examine the actual performance of the pole-mass definition in $\Gamt$, which was presented in ref.~\cite{Chen:2023osm}.
The leading-color part of $\Gamt$ in ref.~\cite{Chen:2023osm}, which accounts for the majority of the total contributions, was checked against the analytical result derived in ref.~\cite{Chen:2023dsi}, and the complete set of $\Oals{3}$ fermionic-loop corrections was derived in ref.~\cite{Fael:2023tcv}.
Based on the complete fixed-order result~\cite{Chen:2023osm}, ref.~\cite{Yan:2024hbz} performed an improved analysis of the total $t$-quark decay width using the principle of maximum conformality scale-setting~\cite{Brodsky:1982gc,Mojaza:2012mf,Brodsky:2012rj,Wu:2014iba} approach. 
In addition, the impact of the $\Oals{3}$ perturbative corrections to the $W$-helicity fractions $\HFs$ as well as the related asymmetries and $W$-energy distribution are firstly addressed in ref.~\cite{Chen:2023osm}. 
~\\

The semi-leptonic decay of a $b$-quark and $c$-quark, namely $\bulv$ and ~$\cqlv$, in perturbative QCD with, respectively, only the decaying heavy quark kept massive involves virtually the same set of Feynman diagrams as the $t$-quark decay, except for having a different number of active quark flavors. 
However, at least in two aspects, additional complexities arise as compared to the $t$-quark case: 
(i) the relatively large value of $\als$ at the mass scale of $c$-quark, $\als(\mcms) \approx 0.38$, which is about twice of $\als(\mbms)$ --- almost twice of $\als(\mtms)$ --- naturally leads to concerns regarding the rate of the convergence of the high-order QCD corrections;  
(ii) $b$-,~$c$-quark and other lighter quarks are confined within hadrons while they decay, and it is thus the matrix elements of the weak currents between hadrons that actually enter the expressions for the physical decay observables, which involve non-perturbative QCD dynamics. 
Both imply that the fixed-order perturbative results alone may not be adequate for making theoretical predictions for the semi-leptonic decays of heavy-flavored mesons that are directly comparable with experiments.  
The Operator Product Expansion (OPE) and Heavy Quark Expansion Theory (HQET)~\cite{Shifman:1984wx,Georgi:1990um,Bigi:1992su,Neubert:1993mb} are among the convenient and powerful theoretical frameworks for describing these decay processes, especially for systematically organizing the non-perturbative contributions. 
The inclusive semi-leptonic decay\footnote{Following the usual convention in the literature on semi-leptonic heavy-meson decays, being ``\textit{inclusive}'' or ``\textit{exclusive}'', here and in the following text, refers to whether or not all hadronic degree-of-freedoms in the final states are summed up, irrespective of whether the final-state leptonic kinematics are kept \textit{differential}.} of B-mesons offers the theoretically cleanest way to determine the CKM mixing angles describing the weak couplings of $b$-quark to W boson, where non-perturbative corrections can be parameterized systematically %using OPE, 
in terms of the hadronic matrix elements of composite operators. 
A direct determination of $|V_{ub}|$ is possible based on the analysis of $\BXulv$ decays, either in the inclusive $\BXulv$ (i.e.~incorporating all final-state hadronic d.o.f.s) or exclusive, e.g.~$\Bpilv$, channels. 
Inclusive decays based on a measurement of the total $\BXulv$ decay rate potentially offer the most accurate way to determine $|V_{ub}|$. 
However, the current results from the inclusive and exclusive determinations disagree significantly~\cite{HFLAV:2019otj,HFLAV:2022esi,ParticleDataGroup:2022pth}. 
The reason for this discrepancy remains unclarified: 
the $|V_{ub}|$ extracted from inclusive $\BXulv$ varies depending on the chosen precision observables and kinematic fiducial regions~\cite{Belle-II:2018jsg,HFLAV:2019otj}; 
inconsistent experimental or theoretical inputs or underestimated systematic errors in the chosen kinematic regions as well as non-SM physics are all among the possibilities~\cite{Belle-II:2018jsg,Gambino:2020jvv}.
Given the critical role of the heavy-to-light structure functions in the inclusive and differential precision observables relevant for extracting $|V_{ub}|$ from inclusive $\BXulv$, the high-order perturbative corrections to these quantities are thus highly desirable, especially in view of the quest to resolve the above confusing discrepancy at Belle II~\cite{Belle-II:2018jsg,Belle-II:2022cgf}.

There is another practical reason, from the experimental side, for obtaining the perturbative corrections not only at higher orders but also remaining differential in leptonic kinematics, e.g.~computing the triple-differential decay rates: 
in the case of charmless semileptonic B-meson decays $\BXulv$, a series of kinematic cuts needs to be applied to isolate the charmless decays, i.e.~to suppress the more abundant $\BXclv$~\cite{Bauer:2000xf,Bauer:2001rc,Belle:2003vfz,Belle:2021ymg}.
It is thus particularly encouraging to note that the first measurements of the differential branching fractions or decay rates of $\BXulv$ were reported in~ref.~\cite{Belle:2021ymg}, very remarkably, as functions of the lepton energy, the four-momentum-transfer squared, light-cone momenta, the hadronic mass, and the hadronic mass squared. 
The measured distributions can be used for novel and more reliable inclusive determinations of $|V_{ub}|$ and the non-perturbative OPE or HQET parameters involved in $\BXulv$ as well as the $b$-quark mass, with reduced dependence on the theoretical models of the non-perturbative dynamics~\cite{Belle:2021ymg}, and eventually for shedding new lights onto the persistent tension with respect to exclusive determinations~\cite{HFLAV:2019otj,HFLAV:2022esi} at Belle II. 
To this end, the knowledge of the high-order perturbative corrections, differential in kinematics, is among the indispensable theoretical ingredients for improving the determination of $|V_{ub}|$ and non-perturbative parameters from inclusive $\BXulv$ measurements with a variety of kinematic cuts~\cite{Bigi:1993fe,Bigi:1993ex,Neubert:1993ch,Neubert:1993um,Bauer:2000xf,Bauer:2001rc,Bosch:2004bt,Bosch:2004th}. 
Given the ambitious goal of measuring inclusive $|V_{ub}|$ at the unprecedented $\mathcal{O}(0.01)$-level at Belle II~\cite{Belle-II:2022cgf}, more precise perturbative corrections to $\BXulv$ are clearly required, e.g.~better than the current 5\% uncertainties for the inclusive decay width~\cite{Hoang:1998hm,Uraltsev:1999rr} and beyond the approximated $\mathcal{O}(\als^2)$~\cite{Gambino:2007rp} for the differential distributions. 
~\\

As mentioned above, the issue of perturbative convergence may be more of concern in application to $\cqlv$ due to the larger $\alpha_s$ at the low energy scales around $c$-quark mass. 
In view of the leading IR-renormalon~\cite{Bigi:1994em,Beneke:1994sw,Beneke:1998ui} in the  pole mass definition and the insight into the possible sources of the large QCD corrections, the choice of the quark mass schemes may have a more pronounced impact here; the behavior of the convergence of the first few perturbative terms may be affected considerably.
Nevertheless, the knowledge of higher-order perturbative QCD corrections is clearly longed-for in a recent comprehensive study~\cite{King:2021xqp} of D-mesons lifetimes, of their ratios and of the inclusive semileptonic decay rates.
Irrespective of whether the knowledge of higher-order perturbative QCD corrections to $\cqlv$ is useful to resolve the issues reported in ref.~\cite{King:2021xqp}, this information may be of help to clarify, in the first place, the question of convergence and hence applicability of the perturbative QCD calculations for this process.
In this regard, it is quite reassuring to know that in refs.~\cite{Shao:2025vhe,Shao:2025qwp} the first simultaneous determination of the CKM matrix elements $V_{cs}$ and $V_{cd}$ together with OPE or HQET non-perturbative parameters was recently performed through a global fit to data from inclusive D-meson decays, where our perturbative QCD corrections for the inclusive lepton-energy moments in $\Dqlv$ up to $\Oals{2}$ were employed and had played an important role in improving the precision of the fit. 
On the other hand, there has been recently very promising progress on the first-principles calculation of $\Dqlv$ from lattice QCD~\cite{Hashimoto:2017wqo,Gambino:2020crt,Gambino:2022dvu,DeSantis:2025qbb,DeSantis:2025yfm,Kellermann:2025pzt}; 
and the theoretical predictions~\cite{DeSantis:2025qbb,DeSantis:2025yfm} for the first few inclusive lepton-energy moments agree with the experimental measurements~\cite{CLEO:2009uah,BESIII:2021duu} within errors, although the total errors of the theoretical results are still larger than the experimental ones~\cite{BESIII:2021duu}.  
~\\

In this work, we are mainly concerned with the high-order perturbative QCD corrections to the heavy-to-light hadronic structure functions or form factors underlying the triple-differential semi-leptonic decay rates~\cite{Manohar:2000dt} of heavy quarks. %, and present the ready-to-use high-precision results up to $\Oals{3}$ for the first time.
These hadronic structure functions encode all strong interaction physics relevant for inclusive semi-leptonic decays, which are perturbative for heavy (free) quarks but contain non-perturbative dynamics for heavy-flavored hadrons. 
Various double and single differential precision observables, as well as inclusive ones, can be composed out of these ingredients, which include the lepton-energy spectrum, the leptonic invariant-mass spectrum, the energy and invariant-mass spectrum of the hadronic system, the double differential decay rates in any pairwise combination of the aforementioned kinematic variables, moments of these distributions with or without cuts on the leptonic kinematics, lepton forward-backward asymmetries e.t.c.
Many of these precision observables have been accurately measured for $\BXclv$ and $\BXulv$ at Belle II ~\cite{Belle:2006kgy,Belle:2021idw,Belle-II:2022evt,Belle:2021ymg,Belle:2023asa} and for $\Dqlv$ at BES III~\cite{BESIII:2021duu,BESIII:2024kvt}, which serve as the key experimental inputs for the ever-more precise extraction of the CKM elements and relevant universal non-perturbative parameters.
In particular, precise calculations of the differential decay rates are essential for precision inclusive measurements of $|V_{ub}|$ in order to separate the signal $\BXulv$ from the much more abundant background $\BXclv$.

Given the central role of these hadronic structure functions for making precision studies of semi-inclusive decays, different theoretical methods~\cite{Shifman:1984wx,Georgi:1990um,Bigi:1992su,Bigi:1993fe,Bigi:1993ex,Neubert:1992hb,Neubert:1993ch,Neubert:1993mb,Neubert:1993um,Mannel:1994pm,Bosch:2004bt,Andersen:2005mj,Aglietti:2006yb,Gambino:2007rp,Aquila:2005hq,Gambino:2006wk,Gambino:2016fdy,Gambino:2022dvu} have been developed with huge efforts to compute them to high accuracy (See, e.g.~refs.~\cite{Manohar:2000dt,Lange:2005yw,Gambino:2020jvv,Fael:2024rys} for overviews).
As far as the perturbation QCD is concerned, it is quite challenging to go beyond the one-loop or $\Oals{}$~\cite{Aquila:2005hq} (NLO) while maintaining the full kinematic dependence, and only the so-called BLM-type\cite{Brodsky:1982gc} and partial $\mathcal{O}(\alpha_s^2\,\beta_0)$ corrections are known analytically or readily available~\cite{Czarnecki:1994pu,Bosch:2004th,Lange:2005yw,Gambino:2007rp,Beneke:2008ei,Asatrian:2008uk,Chen:2018dpt}.
(The color-planar part of the purely virtual $\mathcal{O}(\als^3)$ corrections was derived recently~\cite{Datta:2023otd}.)
The effects of the $\mathcal{O}(\alpha_s^2\,\beta_0)$ perturbative corrections to the triple-differential decay rates turn out to be numerically significant~\cite{Gambino:2007rp}. 
Even though all BLM-type $\mathcal{O}(\alpha_s^{N+1}\,\beta_0^{N})$ corrections to hadronic structure functions for $\BXulv$ were known~\cite{Gambino:2006wk} by dressed gluon exponentiation~\cite{Andersen:2005mj} for quite some time, only until very recently ref.~\cite{Broggio:2026edk} presented an up-to-date extraction of these structure functions at $\mathcal{O}(\als^2)$ via a semi-analytic fit approach, while we are wrapping up our work.
(We kindly note that the Monte-Carlo set-up~\cite{Brucherseifer:2013cu} can be employed to compute the complete $\mathcal{O}(\als^2)$ corrections to $\bulv$ at the fully-differential level.)
On the other hand, the perturbative QCD corrections to various inclusive moments in the semi-leptonic heavy-to-light decay $\bulv$ are typically known to $\Oals{2}$~\cite{vanRitbergen:1999gs,Gambino:2011cq,Finauri:2023kte,Egner:2023kxw,Broggio:2026edk}, and we refer to the recent theoretical reviews~\cite{Gambino:2020jvv,Egner:2024lay,Fael:2024rys} (and references therein) for more detailed information.
In this work, we present the long-awaited complete $\mathcal{O}(\alpha_s)$ and $\mathcal{O}(\alpha_s^2)$ perturbative QCD corrections to all five heavy-to-light structure functions underlying the triple-differential semileptonic decay rates of heavy quarks --- results that have eluded us until now. 
(Some of our key findings have been concisely reported in a companion paper~\cite{companionpaper}; this article provides a comprehensive exposition of the full technical details, along with several supplemental calculations and applications.)
~\\

The remainder of the article is organized as follows.
The next section~\ref{sec:HFFs} is devoted to a detailed presentation of our calculation of the high-order QCD corrections to the structure functions, starting with the kinematics~(\ref{sec:kinematics}) and techniques~(\ref{sec:methods}), followed by the numerical results~(\ref{sec:H2LSFresults}).
In section~\ref{sec:apps}, we focus on applications of these results to the a few selected precision observables, such as the inclusive decay widths~(\ref{sec:Tdecay},~\ref{sec:Bdecay}), the $q^2$-spectrum~(\ref{sec:lpmass}) and lepton-energy moments~(\ref{sec:eem}), in the semi-leptonic decays of the $t$-quark~(\ref{sec:Tdecay}), $b$-quark~(\ref{sec:Bdecay},~\ref{sec:lpmass}) and $c$-quark~(\ref{sec:eem}) that are crucial for extracting the relevant CKM elements and/or non-perturbative parameters from the experiments.
In particular, an interesting technical subtlety encountered in the consistent perturbative reformulation of the differential $q^2$-spectrum up to $\Oals{3}$ in a short-distance or threshold mass scheme is addressed in~subsection~\ref{sec:lpmass_reformula}. 
We conclude in section~\ref{sec:conc}.

\section{Hadronic Structure Functions for Heavy-to-light Decay}
\label{sec:HFFs}

\begin{figure}[htbp]
\begin{center}
\includegraphics[scale=1.0]{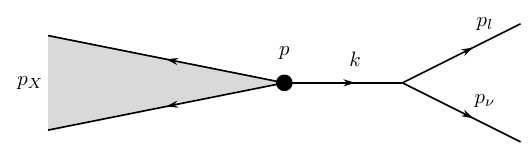}
\caption{The kinematics of the semi-leptonic weak decay of a heavy quark $Q$ with momentum $p$ into a pair of on-shell leptons with momentum $p_l$ and $p_v$, plus a bunch of QCD partons with the total momentum $p_X = p - p_l - p_v$.}
\label{fig:kinematics}
\end{center}
\end{figure}

We begin with describing the kinematics of the semi-leptonic weak decay of a heavy quark $Q$ with momentum $p$ into a pair of on-shell leptons originating from the decay of an intermediate W-boson, the charged one with momentum $p_l$ and the other with momentum $p_v$, plus a bunch of QCD-colored particles with the total momentum $p_X = p - p_l - p_v$. 
This kinematics is illustrated in fig.~\ref{fig:kinematics}.
Since we are concerned with the QCD corrections while working to the leading order in electroweak interaction, we have $p_l + p_v = p - p_X  \equiv k$ and $p_X = p_q + \sum_{i=1}^{n} p_i$ with $p_q$ denoting the momentum of the outgoing quark $q$ that shares the same unbroken fermion line with the decaying heavy quark $Q$ and $p_i$ for the momenta of up to $n=3$ additional QCD partons in the present study. 
Being inclusive in all hadronic d.o.f.s implies that the squared matrix elements will be integrated over all $p_i$ in the partonic phase-space, and eventually only $\{p, p_l, p_v\}$ remain as the independent fixed external momenta.

The central quantity that characterizes the semi-leptonic weak decays of a heavy quark is the heavy-to-light tensor $\mathcal{W}_{Qq}^{\,\mu \nu}$, and in the present work we are concerned with its definition in perturbative QCD: 
\begin{eqnarray} \label{eq:Htensor}
\mathcal{W}_{Qq}^{\,\mu \nu} = 
\frac{1}{\mathrm{N}} \, 
{\textstyle{\sum}} \hspace{-4mm} \int\limits_{X} 
d\Pi_X \, (2 \pi)^4 \delta^4(p - k - p_X) \, 
\langle p |\hat{J}_{Qq}^{\nu +}| p_X \rangle \langle p_X |\hat{J}_{Qq}^{\mu}| p  \rangle \, ,    
\end{eqnarray}
where $ d\Pi_X $ stands for the Lorentz-invariant phase-space integration measure associated with the final state of QCD partons with total momentum $p_X$, and $J_{Qq}^{\mu}$ denotes the heavy-to-light weak-current operator $\bar{\psi}_Q\,\gamma^{\mu}(1-\gamma_5)/2\,\psi_q$.
The conventional initial-state spin and color average factor is collected in $1/\mathrm{N}$. 
The unpolarized inclusive $\HTensor$ depends on two independent momenta $\{p\,,k\}$. 
They can be decomposed into five linearly-independent Lorentz-tensor structures:% 
\begin{eqnarray}
\label{eq:HTffs}    
\mathcal{W}_{Qq}^{\,\mu \nu }(p, k) &=& 
W_1 \,\big( -g^{\mu\nu}\,p^2 \big) \,+ \,
W_2 \, k^{\mu} k^{\nu} + \,
W_3 \, p^{\mu} p^{\nu}  \nonumber\\
&+& W_4 \,  \big(p^{\mu} k^{\nu} \, +\, k^{\mu} p^{\nu} \big) \,-\,
W_5 \, i\epsilon^{\mu \nu \rho \sigma}\, p_\rho \, k_\sigma \, ,
\end{eqnarray}% (each $W_i$ is mass-dimensionless and positive-valued at tree-level for $t \rightarrow b+W$)
where the Levi-Civita tensor $\epsilon^{\mu \nu \rho \sigma}$ appears due to the chiral structure of the weak $tbW$-vertex.\footnote{Since the $\gamma_5$ from the $tbW$-vertex always appears on an open fermion chain of the contributing QCD amplitudes, $\gamma_5$ can be treated fully anticommutively~\cite{Bardeen:1972vi,Chanowitz:1979zu,Gottlieb:1979ix,Korner:1991sx,Korner:1991sx,Kreimer:1993bh,Chen:2024hlv} in a straightforward manner.}
Each form factor $W_i$ is a function of the Lorentz invariants $p^2,\, k^2,\, p\cdot k$ for given $p^2$, and receives both virtual and real-radiation type QCD corrections.
Up to a certain overall normalization factor, these Lorentz quantities are also referred to as the heavy-to-light structure functions that encode all strong interaction physics relevant for inclusive semi-leptonic decays.

The (inclusive) semi-leptonic decay width for a heavy quark can be obtained by the contraction of $\HTensor$ and the leptonic tensor which is the counterpart of \eqref{eq:HTffs} for leptons at the tree-level, up to a factor related to the propagator of the intermediate W-boson. 
In the case of keeping an on-shell W-boson in the decay products, such as in the $t \rightarrow b+W+X_{\mathrm{QCD}}$ for $t$-quark, the differential decay width will be given, up to an overall decay factor, by the projection of $\HTensor$ to the direct product of polarization vectors of the external on-shell W-boson.
More details will be presented later when discussing the numerical results and applications of the hadronic structure functions to the semi-leptonic decays of heavy quarks.

\subsection{Kinematics and phase space}
\label{sec:kinematics}

Being inclusive in all hadronic d.o.f.s implies that the squared matrix elements in \eqref{eq:HTffs} will be integrated over all $p_X = p_q + \sum_{i=1}^{n} p_i$ in the partonic phase-space $d\Pi_X$, and eventually only $\{p, p_l, p_v\}$ remain as the independent fixed external momenta.
Therefore, the full partonic phase-space of the (inclusive) semi-leptonic weak decay of a heavy quark $Q$ can be effectively viewed as a three-particle phase-space but with one variable mass $p_X^2 = \big(p - p_l - p_v \big)^2$.
To facilitate the integration over the hadronic d.o.f.s of the partonic phase-space, we can separate them from the leptonic ones as follows,  
\begin{eqnarray}\label{eq:PSfull}
\mathrm{d}\, \mathrm{PS}_{\mathrm{sl}} 
& \equiv & 
\frac{\mathrm{d}^d\, p_l}{(2 \pi)^d} 2 \pi \delta(p_l^2) \Theta(E_l) \,  
\frac{\mathrm{d}^d\, p_v}{(2 \pi)^d} 2 \pi \delta(p_v^2) \Theta(E_v) \, \nonumber\\
&&\frac{\mathrm{d}^d\, p}{(2 \pi)^d} 2 \pi \delta(p_q^2) \Theta(E_q) \,
\prod_{i=1}^{n} \frac{\mathrm{d}^d\, p_i}{(2 \pi)^d} 2 \pi \delta(p_i^2) \Theta(E_i)     
\, (2 \pi)^d  \delta^d(p - p_l - p_v - p_q - \sum_{i=1}^{n} p_i)  \nonumber\\    
&=& 
\frac{\mathrm{d}^{d-1}\, p_l}{(2 \pi)^{d-1}\, 2\, E_l} \frac{\mathrm{d}^{d-1}\, p_v}{(2 \pi)^{d-1} \, 2\, E_v} \, 
\frac{\mathrm{d}^d\, p_q}{(2 \pi)^d} 2 \pi \delta(p_q^2) \Theta(E_q) \,
\prod_{i=1}^{n} \frac{\mathrm{d}^d\, p_i}{(2 \pi)^d} 2 \pi \delta(p_i^2) \Theta(E_i) 
\, (2 \pi)^d \delta^d(p - k - p_q - \sum_{i=1}^{n} p_i) \nonumber\\
& \equiv & 
\mathrm{d}\, \mathrm{PS}_{\mathrm{L}} \, \mathrm{d}\, \mathrm{PS}_{\mathrm{H}}^{[n]} \,,
\end{eqnarray}
where all leptons and QCD partons in the final states are treated as massless.
Having in mind employing Dimensional Regularization~\cite{tHooft:1972tcz,Bollini:1972ui} to regularize both the intermediate UV and IR divergences, we have kept the explicit dependence on the spacetime dimension $d = 4 -2\varepsilon$. 
More specifically, the dimensionally-regularized hadronic phase-space with $n+1$ QCD partons reads 
\begin{eqnarray}\label{eq:PSH}
\mathrm{d}\, \mathrm{PS}_{\mathrm{H}}^{[n]}
& \equiv & 
\frac{\mathrm{d}^d\, p_q}{(2 \pi)^d} 2 \pi \delta(p_q^2) \Theta(E_q) \,
\prod_{i=1}^{n} \frac{\mathrm{d}^d\, p_i}{(2 \pi)^d} 2 \pi \delta(p_i^2) \Theta(E_i) 
\, (2 \pi)^d \delta^d(p - k - p_q - \sum_{i=1}^{n} p_i) 
\end{eqnarray}
with $k \equiv p_l + p_v$,
and at the lowest order, it is reduced to 
\begin{eqnarray}\label{eq:PSH0}
\mathrm{d}\, \mathrm{PS}_{\mathrm{H}}^{[0]}
&=& 
\frac{\mathrm{d}^d\, p_q}{(2 \pi)^d} 2 \pi \delta(p_q^2) \Theta(E_q) \,
\, (2 \pi)^d \delta^d(p - k - p_q) 
= 2 \pi \delta\big((p - k)^2\big) \Theta(E - E_W)\,.
\end{eqnarray}
The leptonic phase-space $\mathrm{d}\, \mathrm{PS}_{\mathrm{L}}$ can be treated in principle in 4-dimensions, reduced to be 6-dimensional. 
Although one is free to choose the reference frames and axes to parameterize the coordinates of this 6-dimensional integration, the choices with a clear separation between those appearing in the integrand of $\mathrm{d}\, \mathrm{PS}_{\mathrm{L}}$ and those not can greatly simplify the computation.
By virtue of Lorentz invariance, upon the complete inclusive integration over $\mathrm{d}\, \mathrm{PS}_{\mathrm{H}}^{[n]}$, the integrand of $\mathrm{d}\, \mathrm{PS}_{\mathrm{L}}$ can depend at most on the Lorentz invariants made out of $\{p, p_l, p_v\}$.
Taking into account the on-shell kinematic constraints, there are three Lorentz invariants left, which in the rest frame of the decay heavy quark can be parameterized as follows:
\begin{eqnarray}\label{eq:PSLvars}
p \cdot p_l = m_Q \, E_l\,, \quad 
p \cdot p_v = m_Q \, E_v\,, \quad 
p_l \cdot p_v = E_l\, E_v\, \cos \theta_{lv}, \quad 
\end{eqnarray}
where $m_Q$ is the on-shell mass of the decay quark, and $E_l\,, E_v\,$ denote respectively the energies of the two leptons with angle $\theta_{lv}$ between their momentum directions .
Since the absolute orientation of the underlying coordinate reference frame is physically irrelevant ---the integrand does not depend on it---one can always take the momentum $\vec{p}_l$ as the polar axis, and integrate over the spatial angles trivially.
On top of this, only the polar angle $\theta_{lv}$ of $\vec{p}_v$ w.r.t.~$\vec{p}_l$ appears in the integrand, and one can thus integrate over the azimuthal angle of $\vec{p}_v$ trivially.
As a result, one ends up with the following 3-dimensional reduced form of $\mathrm{d}\, \mathrm{PS}_{\mathrm{L}}$ in 4-dimensions:
\begin{eqnarray}\label{eq:PSLred_v0}
\mathrm{d}\, \mathrm{PS}_{L3}
&=& \frac{2\, \pi^2}{ (2 \pi)^6 } |\vec{p}_l|  |\vec{p}_v| \, \mathrm{d} E_l\, \mathrm{d} E_v\, \mathrm{d} \cos \theta_{lv}
\end{eqnarray}
which hold even if the two on-shell leptons are massive, name $m_l^2 = E_l^2 - |\vec{p}_l|^2 >0$ and $m_v^2 = E_v^2 - |\vec{p}_v|^2 >0$.
The integrand of $\mathrm{d}\, \mathrm{PS}_{L3}$ is usually referred to as the triple-differential distribution for the inclusive semi-leptonic decay.
Alternative parameterization of the above 3-dimensional reduced form of the leptonic phase-space can be readily obtained upon a change of the integration variables.

\begin{figure}[htbp]
\begin{center}
\includegraphics[scale=1.0]{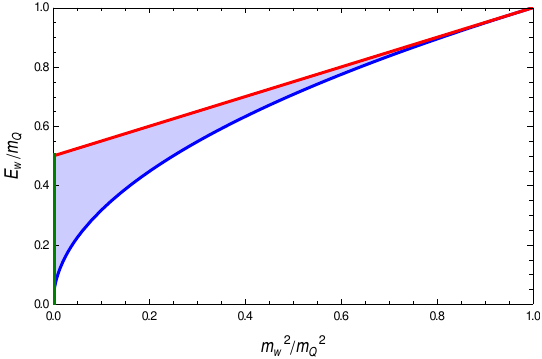}
\caption{The 2-dimensional phase-space in terms of the rescaled kinematic variables $\{m_w^2/m_Q^2,\, E_w/m_Q\}$.}
\label{fig:Ew_mw2_phasespace}
\end{center}
\end{figure}

In view of the fact that the hadronic tensor $\mathcal{W}_{Qq}^{\,\mu \nu }(p, k)$ in \eqref{eq:HTffs} depend only on $p^{\mu}$ and $k^{\mu}$, rather than individually on the leptonic momenta, its Lorentz-invariant form factors or hadronic structure functions shall depend only on $k^2 \equiv m_w^2 \geq 0$ and $k \cdot p = m_Q\, E_w$ with $E_w \equiv E_l + E_v$ (given the fixed $p^2 = m_Q^2$).
Consequently, it is more convenient to reformulate the $\mathrm{d}\, \mathrm{PS}_{L3}$~\eqref{eq:PSLred_v0} in terms of this set of coordinates into the following neat form:  
\begin{eqnarray}\label{eq:PSLred}
\int \mathrm{d}\, \mathrm{PS}_{L3}
&=& \frac{\pi^2}{ (2 \pi)^6 }\, 
\int_{0}^{m_Q^2} \mathrm{d} m_w^2 \, \int_{m_w}^{E_w^{\mathrm{max}}} \mathrm{d} E_w\, \int_{E_l^{\mathrm{min}}}^{E_l^{\mathrm{max}}} \mathrm{d} E_l
\end{eqnarray}
with $E_w^{\mathrm{max}} \equiv (m_Q + m_w^2/m_Q)/2$, $E_l^{\mathrm{max}} \equiv (E_w + \sqrt{E_w^2 - m_w^2})/2 $ and $E_l^{\mathrm{min}} \equiv (E_w - \sqrt{E_w^2 - m_w^2})/2$. 
We will take this parameterization in the following calculation of various differential decay rates.
We have indicated in \eqref{eq:PSLred} the region of the physical phase-space in terms of the integration boundaries of the chosen variables, which can be derived by viewing the kinematics illustrated in fig.~\ref{fig:kinematics} as a sequence of consecutive 1-to-2 splitting processes. 
Figure~\ref{fig:Ew_mw2_phasespace} illustrates the 2-dimensional phase-space in terms of the rescaled coordinates $\{m_w^2/m_Q^2,\, E_w/m_Q\}$. 
Furthermore, with the aid of the linear relations $p_X^2 = (p - k)^2 = m_Q^2 - 2\, m_Q \, E_w + m_w^2$ and $k^2 = (p - p_X)^2 = m_Q^2 - 2\, m_Q \, E_X + p_X^2$, it is straightforward to replace the integration variables in $\int \mathrm{d}\, \mathrm{PS}_{L3}$ by the invariant-mass squared $p_X^2$ and/or total energy $E_X$ of the hardonic system, depending on the differential distributions in question.
For example, another frequently used parametric form facilitating application of cuts in the invariant masses of the leptonic and hadronic systems~\cite{Bauer:2001rc,Belle:2003vfz} reads, 
\begin{eqnarray}\label{eq:PSLred_pw2px2}
\int \mathrm{d}\, \mathrm{PS}_{L3}
&=& \frac{\pi^2}{ (2 \pi)^6 }\, \frac{1}{2\, m_Q}\,
\int_{0}^{m_Q^2} \mathrm{d} m_w^2 \, \int_{0}^{(m_Q - m_w)^2} \mathrm{d} p_X^2\, \int_{E_l^{\mathrm{min}}}^{E_l^{\mathrm{max}}} \mathrm{d} E_l
\end{eqnarray}
but with the \textit{same} $E_l^{\mathrm{min}} $ and $E_l^{\mathrm{max}}$ re-expressed as functions of $m_w$ and $p_X^2$ via $E_w = (m_Q^2 + m_w^2 - p_X^2)/(2\, m_Q)$.
To change $E_w$ to the total energy of the hardonic system, one simply appeals to $E_X = m_Q - E_w$. 
Essentially, there is a simple one-to-one mapping between any pair of $\{m_w^2,\, E_w\}\,, \{m_w^2,\, E_X\}\,, \{m_w^2,\, p_X^2\}\,, \{p_X^2,\, E_X\}$ e.t.c in the 2-dimensional physical phase-space in fig.~\ref{fig:Ew_mw2_phasespace}. 
(The relations among them are linear and independent of $E_l$.)
Below in this article, we stick to the choice $\{m_w^2,\, E_w\}\,$, but it is straightforward to switch to any alternatives.  
To construct the 3-dimensional phase-space in \eqref{eq:PSLred}, for each point in the 2-dimensional area of fig.~\ref{fig:Ew_mw2_phasespace}, one can draw a segment $E_l\, \in \, [E_l^{\mathrm{min}},\, E_l^{\mathrm{max}}]$ in the direction perpendicular to the plane.
Although the form factors $W_i$ depend only on $m_w$ and $E_w$, the triple-differential decay rates do depend on $E_l$, albeit only in the form of a polynomial in the leading-order approximation in electroweak interaction (See~\eqref{eq:3Dparametric} for more details).

It is very illuminating to examine closely the physical configurations corresponding to the boundaries of the 2-dimensional phase-space illustrated in fig.~\ref{fig:Ew_mw2_phasespace}.
\begin{itemize}
\item 
The upper red-colored edge described by the line $E_w/m_Q = (1 + m_w^2/m_Q^2)/2$ corresponds to the kinematic configuration where all additional QCD radiations are either vanishing or collinear to the momentum $p_q$ of the final-state $q$-quark.
Therefore, both the Born-level contribution and all virtual corrections are located here.

\item
The lower blue-colored edge described by $E_w = m_w$ corresponds to the situation with the intermediate W-boson being produced at rest $\vec{p}_w = 0$, namely the two leptons are moving back-to-back with the same energy in the rest frame of the decay process.
The intersection point between the blue and red edge, i.e.~the upper-right node $(1,1)$, corresponds to the extreme case where the total energy of the decaying heavy quark is fully transferred into the leptonic system, which is of course only feasible if all final-state quarks are treated massless.

\item
The vertical green-colored edge, i.e.~the segment of the vertical axis between $0$ and $1/2$, corresponds to the scenario where the two massless leptons move collinearly to each other (in the same direction).
The intersection point $(0,1/2)$ between the green and red edge corresponds to the situation where the leptons and QCD partons form, respectively, two anti-collinear beams moving exactly back-to-back and thus share half of the total energy $m_Q$ of the decaying heavy quark.
On the other hand, the intersection point $(0,0)$ between the green and blue edge corresponds to the case where the total energy  $m_Q$ is entirely transferred into the hadronic system, i.e.~none carried away by the leptons.
\end{itemize}
Lastly, let us remark that one may perform the change of variable: $E_w \rightarrow E'_w \equiv \frac{2\, m_Q^2}{(m_Q - m_w)^2} \big( E_w - m_w \big)$ to turn this 2-dimensional strip area into a rectangular square with the new coordinates $\{m_w^2,\, E'_w \}$. 
This will be employed when we plot the numerical results for the hadronic structure functions in the later subsection~\ref{sec:H2LSFresults}.

\subsection{Method for calculation}
\label{sec:methods}

As introduced earlier in this section, each hadronic structure function $W_i$ is a function of the Lorentz invariants $m_w^2,\, E_w$ for a given $p^2 = m_Q^2$, and receives both virtual and real-radiation type QCD corrections. 
These perturbative QCD corrections are computed in terms of Feynman diagrams, which are generated by the diagram generator~{\fontfamily{qcr}\selectfont
DiaGen}~\cite{diagen} where the Lorentz and Dirac algebra are done using {\fontfamily{qcr}\selectfont
FORM}~\cite{Vermaseren:2000nd}, and in a parallel setup using {\fontfamily{qcr}\selectfont
QGRAF}~\cite{Nogueira:1991ex} and {\fontfamily{qcr}\selectfont
CalcLoop}~\cite{calcloop} for cross-checks.  
It is nowadays standard practice to reduce the loop integrals in these virtual and real-radiation corrections, i.e.~loop integrals without and with on-shell cuts, by the integration-by-parts (IBP) reduction~\cite{Chetyrkin:1981qh,Anastasiou:2002yz} into linear combinations of a much smaller set of the so-called master integrals (MI). 
Without the closed-form results for the complete set of MIs involved in these two-variable functions $W_i$ in the present work, which are currently complete out of reach at $\mathcal{O}(\alpha_s^3)$, we proceed in the following semi-numerical hybrid treatment to get their solutions in the 2-dimension plane $\{m_w^2,\, E_w \}$ illustrated in fig.~\ref{fig:Ew_mw2_phasespace}.

In principle, the system of bivariate partial DEs for MIs in $y \equiv\, m_w^2/m_Q^2$ and $x \equiv E_w/m_Q$ can be derived and solved systematically~\cite{Kotikov:1990kg,Gehrmann:1999as,Caffo:2008aw,Czakon:2008zk}.  
However, for our particular system of $\mathcal{O}(10^3)$ MIs at $\Oals{3}$, we find it practically more accessible and technically efficient to suitably combine the DE method~\cite{Kotikov:1990kg,Gehrmann:1999as} and an efficient interpolation technique into a hybrid treatment.
Our DE-interpolation hybrid strategy can be outlined as follows.  
We first carefully sample a given number $N_p$ of points in the interested regular domain in $y$ according to the piece-wise or stratified Gauss-Kronrod (GK) rules, namely dividing the whole domain into certain number of pieces or segments --- which need not be equal-length --- and within each segment we choose a suitable GK rule of appropriate order (which need not be the same among different segments in general). 
This design, compared to simply take a $N_p$-point GK rule in the whole $y$-domain,  
is to achieve a better overall precision-over-cost ratio, more flexibility and to avoid points too close to intermediate (pseudo) singular or boundary points whose high-precision computations are much more expensive.
Furthermore, the motivation for choosing the sampling points within each segment according to GK rules over, e.g.~a uniform sampling rule, is two-fold at least: 
1) it is extremely efficient when computing integrated results by applying the GK integration rules within each regular segments;
2) it proves to be very beneficial~\cite{Ehrich2000Mastroianni} --- in terms of precision lost --- when one performs a Lagrange-interpolation based on GK nodes to determine the values of the integrand at non-sampling points and to integrate over a sub-domain not exactly coinciding with the pre-partitioned segments. 
(It was shown in ref.~\cite{Ehrich2000Mastroianni} that the Lagrange-interpolation polynomials associated with the interpolation process based on Gauss-Kronrod nodes have the same speed of convergence as the polynomials of best approximation for functions satisfying merely mild conditions.)  
For each sample point in $y$, the MIs viewed as univariate functions in $x$ are then solved using the DE method in terms of piece-wise power-log series (PSE)~\cite{Caffo:2008aw,Czakon:2008zk}, but with the dimensional regulator $\varepsilon$ assigned with numerical values~\cite{Liu:2022mfb,Liu:2022chg} (See below for more comments). %to regularize the intermediate UV and IR divergences (as well as any that can be dimensionally regularized). 
For each requested point $\{x^{0}, y^{0}\}$ in the 2-dimensional phase space, if $y^{0}$ coincides with any of the sampled GK points, the values of the MIs, and any of their linear combinations, can be returned by directly evaluating the known PSE solutions at very low computational cost.
Otherwise, the values will be fitted or extrapolated using an efficient interpolation based on the MIs' values at the set of GK points firstly selected within the very segment where $y^{0}$ belongs to; 
if the precision of the so-interpolated values does not meet the requirement, more sample points from the neighboring segments may be incorporated in the interpolation process, provided there are still enough significant digits left and no singularities are approached by the region enveloping the chosen points. 
It is important to note that the function basis employed in the interpolation or linear-fit ansatz needs not be purely polynomials, but can contain inverse monomials and/or logarithmic factors (if there are singularities sufficiently close to the region in question).\footnote{For example, in our application to $b \rightarrow u \ell \bar{\nu}_{\ell}$ with massive final-state leptons $l = \mu,\, \tau$ to be reported elsewhere, we find that inverse monomials shall be included in the linear ansatz in order to achieve a much more efficient interpolation in the bin containing the lepton-mass-dependent minimal value of $y$. (On the other hand, power-suppressed logarithmic factors are needed for the bin containing the maximal value of $y$.) }
As a final back-up, in case even higher precision is demanded for the MIs at $\{x^{0}, y^{0}\}$ beyond the limit of the above treatments, a uni-variate DE of these MIs in $y$ (with $x$ fixed at $x^0$) needs to be derived, taking MIs' values at the sampled GK point with $y$ nearest to $y^{0}$ as the boundary condition. 
Employing this efficient hybrid strategy for solving $\mathcal{O}(10^3)$ MIs that combines the (Lagrange) interpolation based on stratified Gauss-Kronrod points in $m_w^2$ with the deeply-expanded series in $E_w$, further armed with reduced numerical $\varepsilon$-dependence, we are able to accomplish the extremely challenging task of obtaining high-precision results for all five $W_i$ in the whole 2-dimensional phase space in $\{E_w,\,m_w^2\}$, for the first time, up to $\mathcal{O}(\als^3)$. 
~\\

To be more specific, for the current problem at hand, we first divide the domain $y \in [0,\, 1]$ into 20 segments of equal length (sufficient for the present application).
And for each segment, we choose a sample of points according to the 7-point GK rule.
This means that ideally, e.g.~in case of a regular integrand, applying this integration rule for each small segment in $y$ can lead to an integral result with about 10 significant digits. We kindly note that provided a sufficient integration in $E_w$ around the neighborhood of $E_w^{\mathrm{max}}$, the resulting distribution in $y$ is finite everywhere within the domain $(0,\, 1)$. 
For each of these 140 points in $y$, we solve the MIs as univariate functions in $x $ between $\sqrt{y}$ and $(1 + y)/2$ using the aforementioned DE method in terms of PSE.  
To fix the boundary conditions for these DEs of MIs, we utilize the auxiliary mass flow method~\cite{Liu:2017jxz,Liu:2020kpc,Liu:2021wks,Liu:2022mfb} implemented in {\tt AMFlow}~\cite{Liu:2022chg}.
The phase-space integrals over the momenta $p_i$ of the additional QCD radiations in $\mathrm{d}\, \mathrm{PS}_{\mathrm{H}}^{[n]}$ defined in~\eqref{eq:PSH} are treated in the same manner as loop integrals taking advantage of the reverse unitarity~\cite{Anastasiou:2002yz,Anastasiou:2002qz,Anastasiou:2003yy}.
A number of $\mathcal{O}(10^5)$ integrals in our calculations are reduced to a set of $\mathcal{O}(10^3)$ MIs using the package {\tt Blade}~\cite{Guan:2024byi} based on the strategy of block-triangular form\cite{Liu:2018dmc,Guan:2019bcx}.
In short, the reduction is tackled in a two-step manner: 
we first search for the block-triangular linear relations among target integrals via generating and solving the system of IBP identities based on Laporta's algorithm over a finite field, which typically requires much less numerical samplings as compared to the next step;
we then make full use of the block-triangular relations to efficiently generate a large amount of samplings to reconstruct the IBP-reduction coefficients for the target integrals explicitly. 
This strategy reduces the computational time typically by several times, compared with the reduction without using block-triangular relations. 
We end up with a set of $\mathcal{O}(10^3)$ MIs to be solved using the aforementioned method at each of 140 points in $y$. 
Armed with all these highly efficient techniques, we are able to construct a PSE for each MI, deeply expanded up to about 200 orders in powers of $x$, using the DE solver in {\tt AMFlow}.
By the aforementioned hybrid strategy, we have managed to obtain the high-precision semi-numerical results for all $W_i$, and consequently $\HTensor$, in the whole 2-dimensional phase space up to $\mathcal{O}(\als^3)$. 
~\\

Based on the discussion on kinematics at the end of the previous subsection, it is clear that $W_i$ shall not remain finite at some of the boundaries of the 2-dimensional phase-space in fig.~\ref{fig:Ew_mw2_phasespace}, in particular the red-colored edge where $x$ reaches its maximum value for a given $y$. 
This is expected from the physical picture of the decay process: 
this edge corresponds to the kinematic configuration where all additional QCD radiations either have vanishing energy or become collinear to the direction of the final-state $q$-quark (that couples directly to the decaying heavy $Q$-quark via the weak current). %, which includes all virtual corrections.
Finite physical results are expected only after combining these infrared and/or collinear (IR) singular contributions (located right at this edge) and those spread in a finite neighborhood extending from this edge downwards in $x$ for each given $y$, provided that the intermediate IR singularities distributed among these parts are \textit{consistently} regularized to ensure their proper cancellation when combined together. 
Consequently, when plotting the numerical results for $W_i$ in the 2-dimensional phase space in the next subsection~\ref{sec:H2LSFresults}, we will have to limit ourselves to the regular region of the phase-space keeping a sufficient distance from the above singular red edge.
~\\

On the other hand, to compute the contributions in these IR-singular regions, needed, e.g.~for the inclusive decay widths, moments and the differential distributions with $E_w$ integrated over these regions, we employ the Dimensional Regularization~\cite{tHooft:1972tcz,Bollini:1972ui}(DR) with the dependence on the regulator $\varepsilon = (4-d)/2$ consistently kept in the PSE solutions of all $W_i$. 
To perform the (singular) integration in $E_w$ or $x$ properly, we take the following prescription that is consistent with DR. 
According to the Frobenius series solutions for the differential equations of dimensionally-regularized loop integrals, these integral functions admit the following type of series expansion in any of its kinematic variables, say $x$, around the expansion point parameterized conveniently at $x=0$: 
$$
f(\varepsilon, x) % = \sum_{a,b \in S} x^a \ln^b(x) \, T_{ab}(\varepsilon, x) 
= \sum_{a,b \in S} x^a \ln^b(x) \, \Big(\sum_{n=0}^{\infty} C_{abn}(\varepsilon)\, x^n \Big)
$$
where $a = a_0 + a_1 \varepsilon$ with rational $a_0,\, a_1$ and non-negative integer $b$, belonging to a finite set $S$ (see also,~e.g.,~refs.~\cite{Beneke:1997zp,Smirnov:2004ym,Pak:2010pt,Heinrich:2008si,Lee:2017qql,Liu:2017jxz}). 
We then derive the following term-wise integration formula for $\int_{0}^{u} x^a \ln^b(x) \,\mathrm{d}x$:
\begin{eqnarray} \label{eq:GPLsINT}
%\int_{0}^{u} x^a \ln^b(x) \,\mathrm{d}x  = 
\begin{cases}
\frac{\ln^{1 + b}(u)}{1 + b} & \text{if } a = -1\\
 u^{1+a} \frac{1}{(1+a)^{1+b}} \sum_{i=0}^{b} (-1)^{b-i}\, \frac{b!}{i!}\, (1+a)^{i} \, \ln^{i}(u) & \text{if } a \neq -1
\end{cases}
\end{eqnarray}
where the contribution from $x=0$ is dropped by virtue of DR. 
The phase-space integration of $\HTensor$ over the IR-dangerous regions in $k$ is done using its PSE according to \eqref{eq:GPLsINT} with $\varepsilon$ assigned with the values, $10^{-3} + n \times 10^{-4}$ for $n=0,~1$ (and a few more for inclusive quantities for the sake of intermediate cross-checks), to regularize the potential IR divergences~\cite{Liu:2022mfb,Liu:2022chg}.
The fit regarding the $\varepsilon$-dependence is done only at the very end for the final finite physical objects of interest, which can be the IR-safe inclusive and differential decay widths.
The reduction in computational cost comes mainly from the following point: 
the number of numerical $\varepsilon$-samples needed to fit a precision result merely for the finite objects in the 4-dimensional limit $\varepsilon=0$ (i.e.~only the leading term in their $\varepsilon$-series) is much less than that needed to fully restore all $\varepsilon$-coefficients of the deep Laurent $\varepsilon$-series of the intermediate divergent quantities with the same numerical precision requirement on all $\varepsilon$-coefficients (which involve at least $\mathcal{O}(\varepsilon^0)$).  
In addition, this treatment also saves us from keeping an additional symbol $\varepsilon$ explicitly and performing Laurent expansions in $\varepsilon$ during the intermediate stages, which significantly reduces the symbolic computation time. 
This approach is anticipated to work also with physical jet-based observables, where the dimensionally-regularized phase-space can be conducted using numerical integration methods such as the Monte Carlo technique, with certain phase-space cuts imposed. 
(The numerical values for $\varepsilon$ shall be adapted accordingly.)
Nevertheless, achieving high-precision results for multi-fold differential observables straightforwardly in this way will inevitably lead to escalated computational expenses, highlighting the necessity for further improvements.

In the end our perturbative results for all hadronic structure functions $W_i$ are saved in terms of 140 deeply-expanded PSE per $W_i$ at each order in $\alpha_s$, valid in the whole 2-dimensional physical phase space in $\{m_w^2,\, E_w\}$. To account for the intermediate IR singularities presented in the individual contributions (with even different partonic phase-space), the dependence of these PSE on $\varepsilon$ are retained numerically. More specifically, these PSE are evaluated with $\varepsilon$ assigned with the values $10^{-3} + n \times 10^{-4}$ for $n=0,~1$, and a good strategy for the extrapolation to $\varepsilon \rightarrow 0$ is to have it done only at the very end for the particular finite physical objects of interest.\footnote{Given the size of the complete data, they will be available from the authors only upon request.}
Following from the general theory on extrapolation, we expect at least four significant digits in our final extrapolated results for the $\Oals{3}$ corrections, sufficient for the present phenomenological applications; more precise results can be readily obtained by evaluating additional samples in $\varepsilon$.
This point is confirmed via explicit comparison of the integrated values obtained using these so-extrapolated results against the more accurate values (determined with more samples in $\varepsilon$).
More specifically, to validate the above treatment, we took one sample value in $y$, say $y = 112/517$, and then completed the integration of the differential decay width in $x$; 
a finite result is obtained after combining all contributing pieces. 
We cross-checked this number for the total inclusive result against the one calculated by applying the optical theorem where the integration over $k$ and all $p_i$ can be done directly using the reversed unitarity~\cite{Anastasiou:2002yz,Anastasiou:2002qz,Anastasiou:2003yy}.

\subsection{Results up to $\mathcal{O}(\alpha_s^3)$ in QCD}
\label{sec:H2LSFresults}

Before presenting our numerical results, it is instructive to first have a look at the parametric form of the triple-differential semi-leptonic decay rate in terms of the $W_i$. 
Working with massless leptons in the final state, there are actually only three out of the five $W_i$  defined in~\eqref{eq:HTffs} involved\footnote{We note that the mass effect of final-state charged lepton, e.g.~in $b \rightarrow u \ell \bar{\nu}_{\ell}$ with $l = \mu,\, \tau$, can be easily accounted for by adapting the final-state phase space, and thus the $W_i$ in this work can also be used to study these semi-leptonic decay channels~\cite{Ligeti:2021six}.}:  
\begin{eqnarray} \label{eq:3Dparametric}
\frac{\mathrm{d}^3\,\Gamsl}{\mathcal{N}\, \mathrm{d} m_w^2\, \mathrm{d} E_w\, \mathrm{d} E_l} = 
W_1\,  m_Q^2\, m_w^2 
\,+\, W_3\, m_Q^2\, \big(2\,E_w\,E_l  - 2\, E_l^2 - m_w^2/2 \big) 
\,+\, W_5\, m_Q\, m_w^2\, \big(E_w -2\, E_l \big)\,. \nonumber\\
\end{eqnarray}
Since $W_i$ are independent of $E_l$, with $m_w$ and $E_w$ are used to parameterize the phase-space the dependence of this distribution on $E_l$ takes a polynomial form up to the second power. 
Once integrated over $E_l$, $W_5$ further drops from the resulting 2-dimensional distribution in $m_w$ and $E_w$.
To suppress unnecessary complications, we have pulled out an overall factor $\mathcal{N}$ that collects all necessary factors needed to comply with the standard definition of $\Gamsl$, which includes the initial-state normalization and spin/color-averaging constant factors, as well as a Briet-Wigner propagator factor for the intermediate W-boson (which depends on $m_Q^2$ and $m_w^2$). 
It suffices to specify the following explicit tree-level expressions for $W_i$ in~\eqref{eq:3Dparametric} in four dimensions:
$$W_1 = 12\,(1 - m_w^2/m_Q^2) \,,\, 
\quad W_3 = 48\,,\, \quad W_5 = 24\,,\, $$ 
for a positively charged heavy-quark decay (e.g.~$t \rightarrow b+l+\bar{v}$) without incorporating the phase-space measure~\eqref{eq:PSH0}.
(They agree with the results in ref.~\cite{Aquila:2005hq} apart from an overall normalization convention.)
As $W_1\,, W_3\,, W_5$ depend only on the mass-dimensionless variable $y = m_w^2/m_Q^2$ and $x = E_w/m_Q$, we have employed the freedom of re-scaling the mass-unit to set\footnote{If one considers the semi-leptonic decay of a free $b$-quark in the usual heavy W-boson approximation, upon completing the triple phase-space integration $\mathrm{d} m_w^2\, \mathrm{d} E_w\, \mathrm{d} E_l$ with the so-defined r.h.s.~of \eqref{eq:3Dparametric} at $m_Q=1$, the over constant factor in need reads $\frac{G_F^2\,m_Q^5 \, |V_{Qq}|^2}{192\, \pi^3}$, which is precisely the well-known tree-level expression in Fermi's theory.} 
$m_Q=1$.
Furthermore, the loop corrections to $W_i$ are firstly computed explicitly in a renormalization scheme where the heavy-quark mass $m_Q$ is renormalized in the on-shell scheme and $\als$ is $\MSbar$-renormalized with $n_l$ massless quark flavors. 
The renormalization constants needed can be found in refs.~\cite{Larin:1993tp,vanRitbergen:1997va,Chetyrkin:1999pq,Czakon:2004bu,Melnikov:2000qh}.

\begin{figure}[htbp]
  \begin{subfigure}[b]{0.31\textwidth}
    \centering
    \includegraphics[scale=0.5]{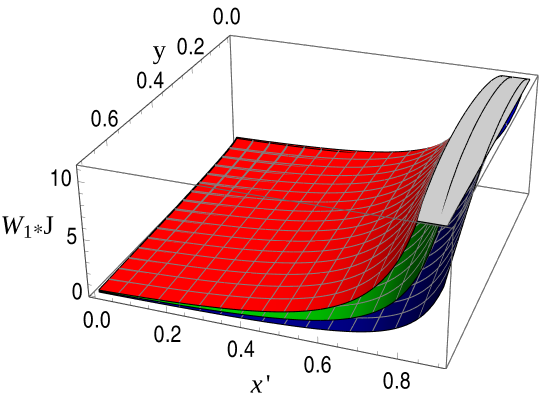}
%    \caption{} \label{fig:diags:1}
  \end{subfigure}
  \begin{subfigure}[b]{0.31\textwidth}
    \centering
    \includegraphics[scale=0.5]{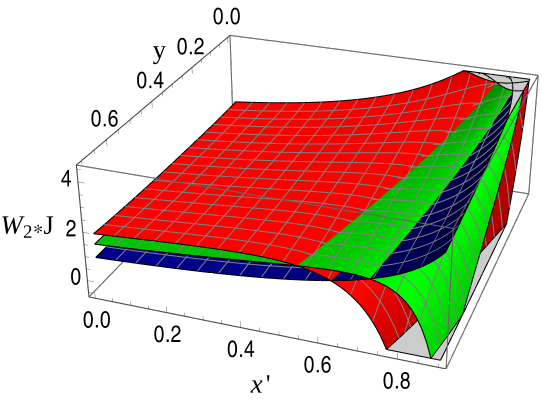}
%    \caption{} \label{fig:diags:2}
  \end{subfigure}%
  \begin{subfigure}[b]{0.31\textwidth}
    \centering
    \includegraphics[scale=0.5]{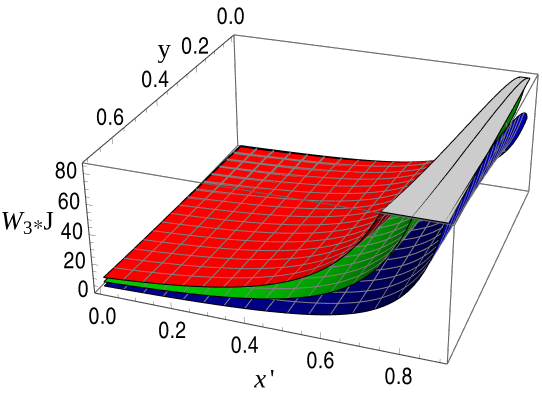}
%    \caption{} \label{fig:diags:3}
  \end{subfigure}%
    \hfill
  \begin{subfigure}[b]{0.05\textwidth}
    \centering
    \includegraphics[scale=0.7]{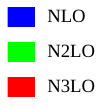}
%    \caption{} \label{fig:diags:1}
  \end{subfigure}%

    \medskip

    \begin{subfigure}[b]{0.31\textwidth}
    \centering
    \includegraphics[scale=0.5]{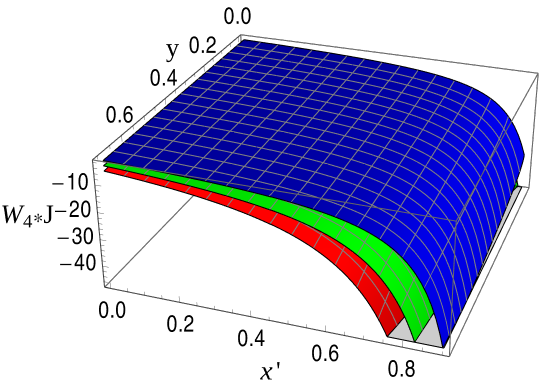}
%    \caption{} \label{fig:diags:4}
  \end{subfigure}%
%  \hfill
  \begin{subfigure}[b]{0.31\textwidth}
    \centering
    \includegraphics[scale=0.5]{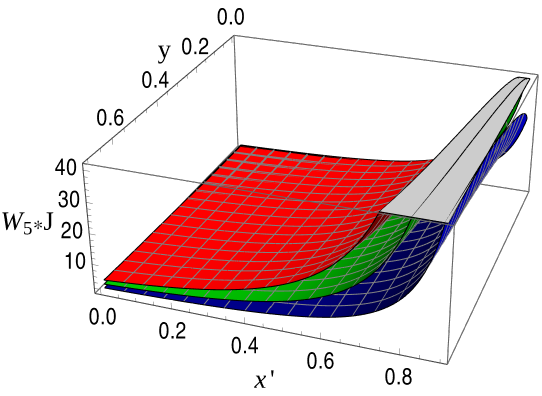}
%    \caption{} \label{fig:diags:5}
  \end{subfigure}%
  \begin{subfigure}[b]{0.31\textwidth}
    \centering
    \includegraphics[scale=0.56]{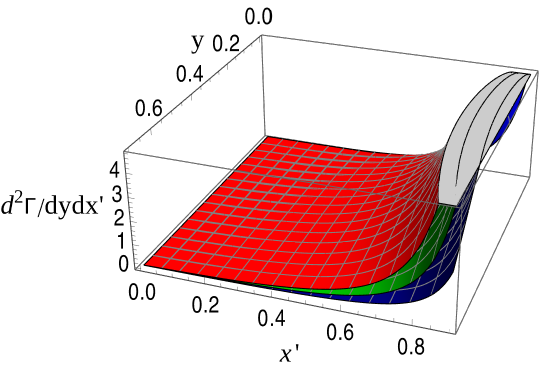}
    %\includegraphics[scale=0.35]{figures/d2Gamma_dmsdEw.pdf}
%    \caption{} \label{fig:diags:6}
  \end{subfigure}%  
    \hfill
    \begin{subfigure}[b]{0.05\textwidth}
    \centering
    \includegraphics[scale=0.7]{figures/Legends_NLO_N2LO_N3LO.pdf}
%    \caption{} \label{fig:diags:1}
  \end{subfigure}%

\caption{
Top row: from left to right are the plots for the numerical results of $W_1\,, W_2\,, W_3$ in a 2-dimensional regular phase-space region $R_2$, multiplied by the Jaccobian of the coordinate transformation, up to $\Oals{1}\,,\Oals{2}\,, \Oals{3}$ indicated respectively by blue, green, and red colors.
Bottom row: from left to right are the plots for the numerical results of $W_4\,, W_5$ and the 2-fold distribution defined in \eqref{eq:2Dparametric} plotted in a similar way as those in the top row.}
\label{fig:WsPlots}
\end{figure}

We are now ready to present the plots for the numerical results of the $W_i$ in a regular region $R_2$ of the 2-dimensional phase-space in figure~\ref{fig:WsPlots}.
We have further inserted $n_l = 4$ and $\alpha_s = 0.22$ for producing these benchmark plots.
The regular phase-space region $R_2$ is selected according to the kinematic constraint $y \in (0, 0.7)$ and $x \in (\sqrt{y}\,,\, 0.9\, x_{\mathrm{max}})$ with $x_{\mathrm{max}} = (1+y)/2$ to avoid the IR-singular edge located at the $x = x_{\mathrm{max}}$. 
For the sake of better illustration, we have changed the coordinates of $R_2$ from the original $\{y,\,x\}$ into $\{y,\,x' = \frac{2\, (x - \sqrt{y})}{(1 - \sqrt{y})^2} \}$, which effectively stretches $R_2$ from the shape of a strip in fig.~\ref{fig:Ew_mw2_phasespace} into a square as indicated by the horizontal plane in fig.~\ref{fig:WsPlots}.
The Jacobian factor $(1 - \sqrt{y})^2/2$ generated by this change of variables is also included in the plots.

In addition to the plots for five structure functions, we have also plotted the result for the  
two-fold differential distribution following from integrating the triple-differential distribution in $E_l$; 
\begin{eqnarray} \label{eq:2Dparametric}
\frac{\mathrm{d}^2\,\Gamsl}{\mathcal{N}\, \mathrm{d} y\, \mathrm{d} x}
= 
\int_{E^{\mathrm{min}}_l}^{E^{\mathrm{min}}_l}\, \frac{\mathrm{d}^3\,\Gamsl}{\mathcal{N}\, \mathrm{d} y\, \mathrm{d} x \, \mathrm{d}\, E_l}\, \mathrm{d}\, E_l
= 
W_1\,  y\,\sqrt{x^2 - y} \,+\, \frac{1}{3}\,\big(x^2 - y \big)^{3/2}\, W_3\,\,  
\end{eqnarray}
in the right-most plot in fig.~\ref{fig:WsPlots} in the same region $R_2$ reformulated using the coordinates $\{y,\,x' = \frac{2\, (x - \sqrt{y})}{(1 - \sqrt{y})^2} \}$. 
Similar generalized ones with additional weight factors $E_l^N$ can be readily composed, the ``doubly-differential'' $N$-th moment of the lepton-energy spectrum~\cite{Gambino:2022dvu}. 
\\

A few comments are in order.
\begin{itemize}
\item 
First of all, the Born kinematics,  where $x = x_{\mathrm{max}}$, is not included in the region $R_2$. 
Consequently, the Born-level result and all virtual-type corrections do not contribute to the plots shown above.
In other words, all contributions covered in $R_2$ must have at least one real QCD-parton with non-vanishing energy and non-collinear momentum, and thus begin at $\mathcal{O}(\alpha_s)$.

\item 
Second, a common feature is clearly exhibited in all these plots: 
the perturbative QCD corrections to these double-differential distributions become larger as $y = m_w^2/m_Q^2$ increases --- namely the invariant mass of the lepton pair becomes larger --- and rise up rapidly as $x = E_w/m_Q$ (or its rescaled counterpart $x'$) gets close to its maximum value.
The latter point is expected: these functions shall not be finite at the boundary of the phase-space where $x$ reaches its maximum value for a given $y$, because this corresponds to the kinematic configuration where all additional QCD radiations are either vanishing or collinear among themselves.   
However, the fixed-order perturbative results need to be combined with suitable resummation~\cite{Andersen:2005mj,Aquila:2005hq,Gambino:2006wk} and/or non-perturbative corrections~\cite{Gambino:2007rp} in order to provide reliable predictions in the neighborhood of these singular boundary regions.

\item 
Lastly, as mentioned at the beginning of this subsection, these perturbative corrections are determined, as usual, with the heavy-quark mass renormalized in the on-shell or pole mass scheme.  
For low-energy processes, such as the $b$- and $c$-quark decay, the convergence behaviors of the first few perturbative terms are already noticeably affected by the so-called leading IR-renormalon issue related to the perturbative definition of the quark pole mass in QCD~\cite{Bigi:1994em,Beneke:1994sw,Beneke:1994rs,Smith:1996xz}.
Improvement in the convergence rate of the perturbative series are typically expected to certain extent, at least regarding the first few terms, after the perturbative corrections to the decay width are properly rewritten\footnote{Strictly speaking, since the dependence on quark mass arise not only from the internal propagators but also from external kinematics, this reformulation entails more than merely a change of the UV mass-renormalization counter-term. Please refer to the discussions in later sections for more details.} in terms of a suitable short-distance quark mass without the leading IR-renormalon issue, such as the commonly used $\MSbar$ mass and many alternative definitions proposed in the literature~\cite{Bigi:1994ga,Bigi:1996si,Czarnecki:1997sz,Fael:2020iea,Beneke:1998rk,Hoang:2008yj,Pineda:2001zq,Komijani:2017vep,Martinelli:1994ty,Aoki:2007xm,Sturm:2009kb,Boyle:2016wis,Chen:2025iul}.

\end{itemize}

\begin{figure}[htbp]
  \begin{subfigure}[b]{0.45\textwidth}
    \centering
    \includegraphics[scale=0.8]{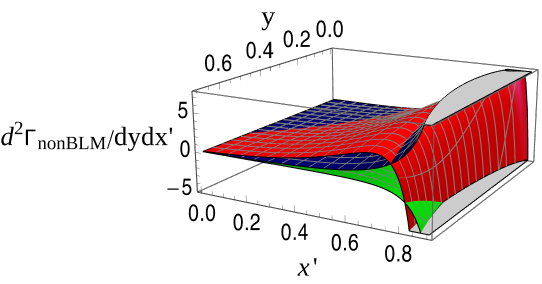}
%    \caption{} \label{fig:diags:4}
  \end{subfigure}%
  \begin{subfigure}[b]{0.05\textwidth}
    \centering
    \includegraphics[scale=0.6]{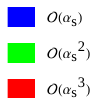}
%    \caption{} \label{fig:diags:1}
  \end{subfigure}%
\begin{subfigure}[b]{0.45\textwidth}
    \centering
    \includegraphics[scale=0.8]{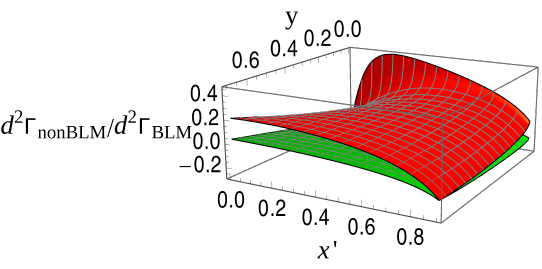}
%    \caption{} \label{fig:diags:4}
  \end{subfigure}%
 \begin{subfigure}[b]{0.05\textwidth}
   \centering
   \includegraphics[scale=0.6]{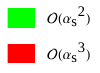}
%   \caption{} \label{fig:diags:1}
 \end{subfigure}%
\caption{The left plot shows the perturbative coefficients after subtracting $\mathcal{O}(\alpha_s^{N+1} \beta_0^{N})$ with $N \geq 1$ at each order in $\alpha_s$, indicated respectively by blue, green, and red colors; the $\Oals{1}$-coefficient is included (in blue color) for reference. 
The right plot shows the ratios of the non-BLM to the BLM-type contribution defined in the decomposition~\eqref{eq:dGam2fold_BLMseperation} at, respectively, $\Oals{2}$ and $\Oals{3}$.}
\label{fig:BLMPlots}
\end{figure}

As is well-known in perturbative QCD, the effective scale of the running QCD coupling for a typical scattering process is related to certain ``gluon virtuality''~\cite{Brodsky:1982gc} which depends on kinematics in general.
Based on the physical picture of the inclusive semi-leptonic decay process, in the kinematic region with small invariant mass for the hadronic system $\,p_X^2 = m_Q^2 + m_w^2 - 2\, m_Q\, E_w $, the effective scale of the QCD coupling shall be typically lower, or even much lower, than the mass of the decay quark (which is usually taken as the default renormalization point). 
Consequently, the so-called running-coupling or Brodsky-Lepage-Mackenzie (BLM)~\cite{Brodsky:1982gc} type corrections to $\bulv$, characterized by the power factor $\mathcal{O}(\alpha_s^{N+1} \beta_0^{N})$, are expected to be large~\cite{Andersen:2005mj,Aquila:2005hq,Gambino:2006wk}, which are related to the leading infrared renormalon~\cite{Bigi:1994em,Beneke:1994sw} of the quark pole mass.\footnote{It has been pointed out recently~\cite{Chen:2025iul,Chen:2025zfa} that the leading IR-renormalon divergence in the pole mass of a massive quark~\cite{Bigi:1994em,Beneke:1994sw} resides entirely in the contribution from the trace anomaly~\cite{Adler:1976zt,Collins:1976yq,Nielsen:1977sy} of the energy-momentum tensor in QCD.}
Indeed, the $\mathcal{O}(\alpha_s^{N+1} \beta_0^{N})$ corrections for $\bulv$ dominate over the remaining non-BLM-type corrections to the inclusive decay width at $\Oals{2}$~\cite{vanRitbergen:1999gs,Hoang:1998hm,Uraltsev:1999rr}. 
For the $W_i$ and the triple-differential decay rates, the $\mathcal{O}(\alpha_s^2 \beta_0)$ BLM-type corrections are expected to account for the major part of the $\Oals{2}$ corrections~\cite{Gambino:2006wk,Gambino:2007rp,Brucherseifer:2013cu} in the on-shell quark-mass scheme, although the exact ratios depend on the kinematic regions.
Here we confirm this expectation for the role of the BLM-type corrections at $\Oals{2}$ in the on-shell scheme, and further extend the analysis to $\Oals{3}$. 
To this end, we decompose the total perturbative results for the double differential decay rate~\eqref{eq:2Dparametric} as follows: 
\begin{equation}\label{eq:dGam2fold_BLMseperation}
\frac{\mathrm{d}^2\,\Gamsl}{\mathcal{N}\, \mathrm{d} y\, \mathrm{d} x} 
= 
\frac{1}{\mathcal{N}} \, \Big(
\frac{\mathrm{d}^2\,\Gamsl_{\mathrm{BLM}}}{ \mathrm{d} y\, \mathrm{d} x} + \frac{\mathrm{d}^2\,\Gamsl_{\mathrm{nonBLM}}}{\mathrm{d} y\, \mathrm{d} x} 
\Big)    
\end{equation}
where each of these two parts admits, respectively, a perturbative expansion in $\als$.  
In our definition~\eqref{eq:dGam2fold_BLMseperation} the BLM-part $\mathrm{d}^2\,\Gamsl_{\mathrm{BLM}} $ collects all $\mathcal{O}(\alpha_s^{N+1} \beta_0^{N})$ corrections for $N \geq 0$ (i.e.~including the full $\Oals{1}$ correction without $\beta_0$), thus the remaining so-defined non-BLM part starts only from $\Oals{2}$. 
In fig.~\ref{fig:BLMPlots} we plot our on-shell perturbative results for the double differential decay rate~\eqref{eq:dGam2fold_BLMseperation} in the regular kinematic region with the BLM and non-BLM type  corrections separated up to $\Oals{3}$.
A general trend exhibited in the ratio plot in fig.~\ref{fig:BLMPlots} is that for a given $q^2 = y\, m_Q^2$ in most of the covered regular region, the larger $E_w$ or $x'$ --- implying a smaller $p_X^2$ of the hadronic system --- the BLM-type contributions dominate over the remaining non-BLM-type contributions, holding both at $\Oals{2}$ and $\Oals{3}$. 
After subtracting the $\mathcal{O}(\alpha_s^{N+1} \beta_0^{N})$ terms with $N \geq 1$, the remaining perturbative contributions exhibit a much better behaved convergence behavior up to $\Oals{3}$ in the regular region as shown by the left plot (with $n_l =4$).
Therefore, we expect that a significantly improved prediction from the perturbation theory for $\bulv$ (and $\cqlv$) may result from carefully merging our fixed-order on-shell results for $W_i$ with the BLM-resummed result~\cite{Gambino:2006wk}, after taking into account properly the intermediate subtraction terms in the whole phase-space, which we leave for near-future work. 

Additional improvements are typically expected once reformulating the quark-mass dependence of the perturbative corrections using short-distance or threshold mass schemes, which can be much more conveniently done for less exclusive or differential observables of interest, obtained via integrating eq.~\eqref{eq:3Dparametric} over appropriate quark-mass-independent kinematic variables. 
We will demonstrate this explicitly below in a few selected applications of these perturbative hadronic structure functions, and comment on an interesting point firstly observed along the way for differential precision observables.

\section{Applications to Semi-leptonic Decays of Heavy Quarks}
\label{sec:apps}

With the perturbative results of the full set of $W_i(m_w^2, E_w)$ becoming available, we can compute various differential and inclusive decay rates by integrating the triple-differential distribution $\frac{\mathrm{d}^3\,\Gamsl}{\mathcal{N}\, \mathrm{d} m_w^2\, \mathrm{d} E_w\, \mathrm{d} E_l}$~\eqref{eq:3Dparametric} over some or all of the kinematic variables $m_w\,, E_w\,, E_l$, possibly with kinematic weight factors included, such as in the definition of moments. 
An example of the double-differential distributions was already composed in~\eqref{eq:2Dparametric} via integrating out $E_l$, and similar generalized ones with additional weight factors $E_l^N$ can be readily composed, the ``doubly-differential'' $N$-th moment of the lepton-energy spectrum~\cite{Gambino:2022dvu}
It shall be then straightforward to transform the differential variables to derive the differential distributions involving the hadronic kinematic variables, such as the invariant mass squared $p_X^2 =  m_Q^2 + m_w^2 - 2 \,m_Q\, E_w $ in \eqref{eq:PSLred_pw2px2} and the total energy of the whole hadronic system $E_X = m_Q - E_w$. 
Due to different energy scales and physical observables involved in the applications to the semi-leptonic weak decays of $t$-, $b$-~and $c$-quark, the best practices to treat the perturbative QCD corrections vary, and it is thus more convenient to discuss different scenarios of the application separately.
Below we discuss in detail a few selected precision observables that are actively studied experimentally.

\subsection{$\Gamt$ for $t \rightarrow b\, W$}
\label{sec:Tdecay}

Let us first consider the semi-leptonic decay of $t$-quark,  $t \rightarrow b + W^*(\ell \bar{\nu}_{\ell})$.
Since $m_t$ is much larger than $\Lambda_{QCD}$ and $\alpha_s(\mu=m_t)$ is quite small, the perturbative results of the hadronc functions alone are already quite sufficient to provide reliable theoretical predictions for this decay process to compare with the experimental measurements. 
We begin with the simplest observable, the total inclusive decay width $\Gamt$. 
As $m_t$ is larger than the physical mass $m_W$ of the W-boson, the latter can be on- and off-shell in $t \rightarrow b + W^*(\ell \bar{\nu}_{\ell})$.
Since the Breit-Wigner propagator of the intermediate W-boson does not depend on $E_l$ and $E_w$, we can easily integrate over these variables, and end up with the following parameteric form: 
\begin{eqnarray}\label{eq:GammaTsl}
\GamTsl &\equiv& 
\int^{m_t^2}_{0} \frac{\mathrm{d} m_w^2}{2 \pi} 
\frac{2\, m_W \, \Gamma_{W \rightarrow \ell \bar{\nu}_{\ell}} }{ 
\big(m_w^2 - m_W^2 \big)^2 + \big(m_W \Gamma_W^{tot} \big)^2} 
\, \tilde{\Gamma}_{t}(m^2_t, m_w^2)  
\end{eqnarray}
where $\Gamma_W^{tot}$ is the total decay width of the W-boson.
The leading-order $\Gamma_{W \rightarrow \ell \bar{\nu}_{\ell}}$ in the numerator is introduced via rewritting the electroweak coupling factors in the W-boson decay vertex: $\frac{g_2^2}{24 \pi} = \frac{2 \Gamma_{W \rightarrow \ell \bar{\nu}_{\ell}}}{m_W} = \frac{m_W^2 \, G_F}{3 \sqrt{2} \pi}$. 
In the narrow-width approximation of the intermediate W-boson, namely $\Gamma_W^{tot} \ll m_W$, 
\begin{eqnarray}\label{eq:GammaTos}
\GamTsl \big|_{\Gamma_W^{tot} \ll m_W} \, \approx \, \tilde{\Gamma}_{t}(m^2_t, m_w^2=m_W^2) \, \frac{\Gamma_{W \rightarrow \ell \bar{\nu}_{\ell}}}{\Gamma_W^{tot}} \,  \equiv\,  \Gamt \, \frac{\Gamma_{W \rightarrow \ell \bar{\nu}_{\ell}}}{\Gamma_W^{tot}}
\end{eqnarray}
where the notation $\Gamt$ is introduced to denote the inclusive decay width of the $t$-quark to an on-shell W-boson and a $b$-quark plus additional QCD radiations, the object of concern in the following text.

\subsubsection{Results in the pole mass scheme}
\label{sec:Tdecay_OS}

We now discuss the numerical results for the on-shell decay width $\Gamt$ introduced in \eqref{eq:GammaTos}.\footnote{The results in the on-shell $t$-mass scheme discussed in this subsection was presented initially in our earlier work~\cite{Chen:2023osm}, but included here both for completeness and contrast with the new results derived in alternative mass schemes.}
The QCD effects on $\Gamt$ in SM can be parameterized as%
\begin{equation}
\label{eq:parametricDW}
\Gamt = \Gamma_0\, \Big[ \mathbf{c}_0 + \frac{\als}{\pi}  \mathbf{c}_1 + \Big(\frac{\als}{\pi}\Big)^2 \mathbf{c}_2 + \Big(\frac{\als}{\pi}\Big)^3  \mathbf{c}_3 
+ \mathcal{O}(\als^4) \Big] \, ,
\end{equation}%
where we introduced a prefactor %$\Gamma_0\equiv\frac{G_F\, m_t^3\, |V_{tb}|^2}{8 \sqrt{2} \pi }$ 
$\Gamma_0\equiv\frac{G_F\,\, m_W^2\, m_t \, |V_{tb}|^2}{12 \sqrt{2} }$ 
with $G_F$ the Fermi constant, the CKM matrix element $V_{tb}$ taken to be 1 in the following numerical results.
The perturbative coefficients $\mathbf{c}_i$ are functions of the kinematic ratio $m_W^2/m_t^2$ in the limit $m_b = 0$, and also the renormalization scale $\mu$ in general.

We first present the numerical results in the renormalization scheme where $m_t$ is renormalized in the on-shell scheme and $\als$ is $\MSbar$-renormalized with 5 massless quark flavors. 
Motivated by the kinetic energy $m_t - m_W - m_b$ of the final state, we provide our benchmark values for the on-shell results at the scale $\mu = m_t/2$, which read %
\begin{eqnarray}
\label{eq:ciV}
\mathbf{c}_0 = 1.93851\,,  \quad 
\mathbf{c}_1 = -4.85519\,, \quad 
\mathbf{c}_2 = -21.2260 \,, \quad
\mathbf{c}_3 = -174.265\,, 
\end{eqnarray}%
where we used the SM input parameters $m_t = 172.69$ GeV and $m_W = 80.377$ GeV.\footnote{For reference, $\Gamma_t(\mu=m_t) = 1.48643\,-\, 0.12751\,-\,0.030697 \,-\, 0.01040 = 1.31782$ GeV, where the first equality provides the decomposition of the total N3LO result according to the $\als$ order.}  
In \eqref{eq:ciV} we have truncated our internal high-precision results for the perturbative coefficients $\mathbf{c}_i$ to the first 6 digits for the sake of display (Results with higher precision can be obtained from the supplementary file.)
Up to N2LO our result~\eqref{eq:ciV} agrees with those in refs.~\cite{Blokland:2004ye,Blokland:2005vq,Brucherseifer:2013iv,Chen:2022wit}.
The leading-color part of $\Gamt$ at N3LO agrees with ref.~\cite{Chen:2023dsi}. 
With $G_F = 1.166379\times 10^{-5} $ GeV$^{-2}$ and $\alpha_s(m_t/2) \approx 0.1189$ at scale $\mu=m_t/2$, obtained by solving the renormalization-group equation for the running $\als$ at four-loop order~\cite{vanRitbergen:1997va,Chetyrkin:1999pq,Czakon:2004bu,Chetyrkin:2004mf} with input $\alpha_s(m_Z) = 0.1179$ at the Z-pole mass $m_Z = 91.1876$ GeV, we further obtain%
\begin{eqnarray}
\label{eq:explicitDW}
\Gamt(\mu=m_t/2) &=& 1.4864 -0.14088 -0.02331 -0.00724 \,\,\text{GeV}  \nonumber\\
  &=& 1.3150\,\,\text{GeV} \, ,   
\end{eqnarray}
where the first line provides the decomposition of the total N3LO result according to the $\als$ order.\footnote{The contributions with closed top-loop diagrams start from two-loop order, which are not included here. We have checked that this part amounts to about $0.2\%$ of the pure $\mathcal{O}(\alpha_s^2)$ correction and thus contributes less than $10^{-4}$ of the total result.} 
Therefore, the QCD corrections continue to decrease the Born-level result for $\Gamt$ up to $\mathcal{O}(\als^3)$. 
Results for $\Gamt$ at other scales can be readily derived from \eqref{eq:ciV} using the renormalization-group equation method. 
~\\

It is appropriate here to briefly comment on a few miscellaneous effects in the theoretical result for the $t$-quark decay width, which was initially discussed in our earlier work~\cite{Chen:2023osm}.
The complete expression for the semi-leptonic decay width $\GamTsl$ is defined in~\eqref{eq:GammaTsl}, with the off-shell-$W$ effect encoded through a Breit-Wigner distribution. 
The perturbative QCD corrections to $\GamTsl$ assume a similar parametrization as in~\eqref{eq:parametricDW} but with the $i$-th order perturbative coefficient denoted as $\tilde{\mathbf{c}}_i$ for distinction.  
With the expression for $\Gamt$ as a deeply-expanded PSE w.r.t. $m_w^2/m_t^2$ at our hand, valid in the whole physical region, we can investigate the off-shell-$W$ effect on $\GamTsl$ up to $\mathcal{O}(\as^3)$.
With $W$ total decay width $\Gamma_W = 2.085$ GeV, we find that $\delta_i \equiv\big(\tilde{\mathbf{c}}_i-\mathbf{c}_i\big)/\mathbf{c}_i$ are quite small and decrease, albeit very slowly, as the $\as$-order increases: 
$\delta_i$ takes $-1.54\%, -1.53\%, -1.39\%, -1.23\% $ respectively for $i=0,1,2,3$.

For the finite $b$-quark mass effect, we denote the $i$-th order coefficient for $\Gamt$ with a non-zero $m_b$ as $\mathbf{c}^{m_b}_i$. 
With $m_b = 4.78\,$GeV we find that $\big(\mathbf{c}^{m_b}_1-\mathbf{c}_1\big)/\mathbf{c}_1 \approx \big(\mathbf{c}^{m_b}_2-\mathbf{c}_2\big)/\mathbf{c}_2 \approx -1.47 \%$.
This strongly indicates that the small non-zero $m_b$ effect at $\mathcal{O}(\as^3)$ may observe a similar small ratio, well-below sub-per-mille level for the total $\Gamt$.
On top of this, there is also the NLO electroweak corrections~\cite{Denner:1990cpz,Denner:1990ns,Eilam:1991iz,Do:2002ky,Basso:2015gca} which may be included by a multiplicative $K$-factor.
We have re-calculated NLO electroweak $K$-factor to be $1.0168$, which agrees with the result in ref.~\cite{Denner:1990cpz,Denner:1990ns,Eilam:1991iz,Do:2002ky,Basso:2015gca}.
It is very interesting to note that if both the NLO electroweak correction and the aforementioned finite $m_b$ and off-shell $W$ effects are taken into account in the \textit{multiplicative} manner, i.e.~in terms of K-factors multiplied onto our pure QCD result~\eqref{eq:explicitDW} as a whole (with on-shell W-boson and massless $b$-quark), 
these two K-factors nearly completely compensate each other, leading to a total K-factor that coincides with 1 up to the first 3 digits (i.e.~at the sub-per-mille level).
In view of this, we will consider in the following discussions only the pure QCD corrections up to $\Oals{3}$.

\subsubsection{Results in short-distance mass schemes}
\label{sec:Tdecay_TS}

Despite an overall decent convergence in \eqref{eq:explicitDW} at first sight, by examining the successive ratios $\mathbf{c}_{i+1}/\mathbf{c}_i$ starting from $i=1$, we begin to see the hint that the convergence of the perturbative series seems to deteriorate as the perturbative order goes higher.
This is expected due to the leading infrared-renormalon~\cite{Bigi:1994em,Beneke:1994sw,Beneke:1998ui}  sensitivity of the on-shell (pole) mass definition for heavy quarks (see, e.g.~refs.\cite{Beneke:1994qe,Smith:1994id,Ball:1995wa,Smith:1996xz,Beneke:1998ui,Beneke:2016cbu,Hoang:2017btd,FerrarioRavasio:2018ubr}). 
To investigate this, we isolated the leading $\alpha_s^n\,\beta_0^{n-1}$-enhanced terms for $n=2,3$ in the perturbative corrections to $\Gamma_t$ in the on-shell mass scheme 
in the so-called large-$n_f$ approximation (or ``naive non-abelianization'')~\cite{Beneke:1994qe,Broadhurst:1994se}. 
We have found that the so-identified coefficients of $\alpha_s^2\,\beta_0$ and $\alpha_s^3\,\beta_0^2$ are very close to those predicted in ref.~\cite{Ball:1995wa}.
Furthermore, we confirm that these terms do dominate over the remaining pieces at each perturbative order, very similarly as in the semi-leptonic $b$-quark decay (to be discussed later). 
We thus conclude that the superficial convergence of the perturbative corrections in~\eqref{eq:explicitDW} for $\Gamma_t$ is mainly owing to the smallness of $\alpha_s(\mu=m_t)$, c.f.~about half of $\alpha_s(\mu=m_b)$. 
Consequently, the ideas developed and proved useful in improving the convergence of the perturbative series for the semi-leptonic $b$-quark, notably by changing into an appropriate short-distance mass scheme, can also be applied to further improve the situation of the $t$-quark decay.

\begin{figure}[htbp]
\begin{center}
\includegraphics[scale=0.8]{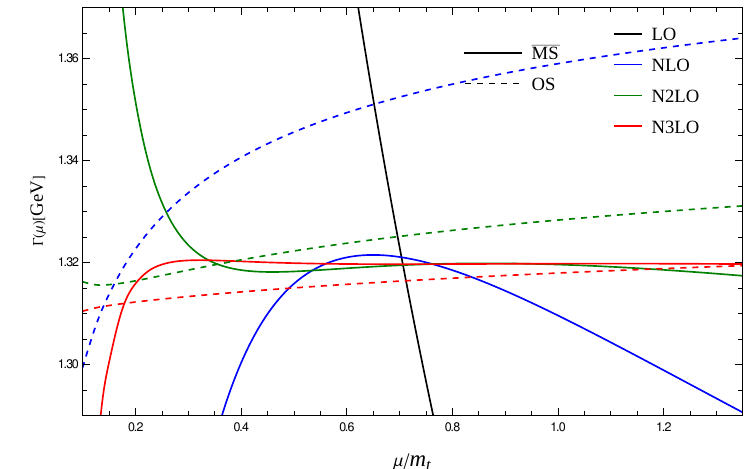}
\caption{The scale dependence of the fixed-order results for $\Gamt$ and $\bar{\Gamma}_t$ in $\mu/m_t \in [0.1, 1.3]$}
\label{fig:TotWidthMuD}
\end{center}
\end{figure}

Using the three-loop conversion formula from the pole mass $m_t$ to $\overline{m}_t$~\cite{Melnikov:2000qh} and decoupling relations\cite{Chetyrkin:1997sg,Chetyrkin:1997un}, we get $\overline{m}_t(\overline{m}_t) = 163.15$ GeV in 5-flavor scheme. 
Then we obtain the results 
\begin{eqnarray}
\label{eq:explicitDWMS}
\bar{\Gamma}_t(\mu=\overline{m}_t/2) = 
1.48847\,-\,0.17561\,+\,0.00528\,+\,0.00180 = 1.3199(12)\, \text{GeV}, 
\end{eqnarray}
where the behaviors of the high-order corrections are improved as compared to \eqref{eq:explicitDW}.
The number in the bracket right after the total value in \eqref{eq:explicitDWMS} corresponds to the change of the last two digits when $\mu$ is varied between $\overline{m}_t/4$ and $\overline{m}_t$.
We note that for $\mu/\overline{m}_t \in [1/2, 1]$, the pure third-order QCD correction in the 5-flavor $\MSbar$ scheme is below one per-mile of the N2LO result.
The pure third-order QCD correction in the 5-flavor $\MSbar$ scheme is accidentally close to zero around $\mu=\overline{m}_t$;
The values of $\bar{\Gamma}_t$ at both $\overline{m}_t/4$ and $\overline{m}_t$ are less than $\bar{\Gamma}_t(\mu=\overline{m}_t/2)$, which is among the reasons why the error in \eqref{eq:explicitDWMS} is very small. 
(In the 6-flavor scheme, this no longer happens and the conventional scale uncertainty doubles.) 
It is very illuminating to examine the scale dependence of our results for $\Gamt$ and $\bar{\Gamma}_t$ in $\mu/m_t \in [0.1, 1.3]$ up to N3LO, which are plotted in fig.~\ref{fig:TotWidthMuD}.
(We have used the renormalization group equations at four-loop order for the running of both $\alpha_s$ and $\MSbar$-mass $\overline{m}_t$ in the 5-flavor scheme in these plots.\footnote{Taking the results in $\MSbar$ as an example, the differences produced between the $\alpha_s$-running prescription 1) using $n$-loop $\als$-running when evaluating the perturbative result truncated up to $\alpha_s^n$-order for $n=1,2,3$  and prescription 2) always using the four-loop $\alpha_s$-running --- which is adopted in making the scale-variations in all shown plots --- are generally small: the change in the size of the conventional scale bands is usually at the level of 1 to 10\% in relative; the higher the perturbative order in question, the less difference between these two $\als$-running prescriptions. }
As expected, the scale dependence are improved for the perturbative results in both mass schemes as higher order corrections are included.
However, in the case of the on-shell results $\Gamt$, 
due to the turning point of the N2LO green solid curve (at about $\mu/m_t=0.14$) in fig.~\ref{fig:TotWidthMuD}, its scale variation can never cover the N3LO result at any scales less than $\mu/m_t=0.6$, including our central scale $\mu/m_t=0.5$.
This clearly demonstrates that the N2LO results in the on-shell scheme can underestimate the theoretical error by simply studying the $\mu$ dependence, and thus determining the $\mathcal{O}(\als^3)$ corrections explicitly is very important. 
We note that at the benchmark scale $\mu = m_t$ typically chosen in literature, the pure $\mathcal{O}(\als^3)$ correction decreases $\Gamt$ by about $0.8\%$ of the previous N2LO result, roughly $10\,$MeV, significantly exceeding the usual N2LO scale uncertainty.
On the other hand, the results for $\bar{\Gamma}_t$ at N2LO and N3LO clearly show improved convergence and reduced scale uncertainties compared to the corresponding results for $\Gamt$, in the bulk of $\mu$-range plotted, as clearly visible in fig.~\ref{fig:TotWidthMuD}.
~\\

\begin{figure}[htbp]
\begin{center}
\includegraphics[scale=0.8]{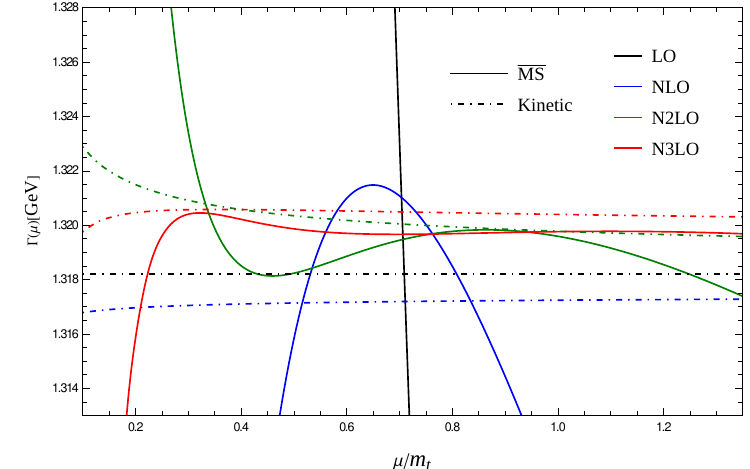}
\caption{The scale dependence of the fixed-order results for $\Gamma_t^{\mathrm{kin}}$ in $\mu/m_t \in [0.1, 1.3]$, and $\bar{\Gamma}_t$ included for comparison.
(Note that the resolution of the y-axis is different from fig.~\ref{fig:TotWidthMuD}.)}
\label{fig:TotWidthMuD_MSvsKin}
\end{center}
\end{figure}

The kinetic mass~\cite{Bigi:1994ga,Bigi:1996si,Czarnecki:1997sz} is a short-distance threshold quark mass specifically designed and well-suited for studying perturbative corrections to semi-leptonic $b$-quark decay. 
Due to the similarity in the structures of the perturbative corrections to $\Gamma_t$ as those in the semi-leptonic $b$-quark decay, it is interesting and instructive to try out the kinetic-mass scheme for $\Gamma_t$, namely express $\Gamma_t$ as a function of the kinetic mass $\mtkin$ of the $t$-quark.
The main difference to be noted here is that the leading power-scaling behavior of $\Gamma_t$ in $m_t$ is 3, rather than 5 as in the semi-leptonic $b$-quark decay~\cite{Bigi:1996si}. 
To this end, the three-loop relation between the pole mass and kinetic mass~\cite{Fael:2020iea} is employed.
We have chosen the so-called Wilsonian cutoff scale~\cite{Bigi:1996si} $\mu_{c} = m_t/3$ for the kinetic mass of $t$-quark, because of the leading $m_t^3$ power-scaling behavior of $\Gamma_t$, and obtained $\mtkin(\mu_{c} = m_t/3) = 166.92$~GeV from $m_t = 172.69$~GeV using the three-loop conversion formula~\cite{Fael:2020iea} (which is very stable against the scale $\mu$ of the conversion). 
Our result for $\Gamma_t$ in the kinetic mass scheme at the $\als$-renormalization scale $\mu = \mtkin/2$ reads: 
\begin{equation}\label{eq:explicitDWKin}
\Gamma_t^{\mathrm{kin}}(\mu = \mtkin/2\,,\, \mu_{c} = m_t/3) 
= 1.31821\,-\,0.00108 \,+\,0.00325 \,+\, 0.00018 = 1.3206(22)~ \text{GeV}
\end{equation}
The scale dependence of our results are shown in fig.~\ref{fig:TotWidthMuD_MSvsKin}, where the results $\bar{\Gamma}_t$ in 5-flavor $\MSbar$ scheme are plotted for comparison; 
and we find that both the scale dependence and convergence of the perturbative series are significantly improved.  
Within a wide range $\mu \in [\mtkin/4, 2\,\mtkin]$, the variation of the N3LO result for $\Gamma_t^{\mathrm{kin}}(\mu_{c})$ with $\mu_{c}$ varied within $[\mtkin/5,\,\mtkin/3]$ is well below one per-mile.
Only when the range is extended to $\mu_{c} \in [m_t^{\mathrm{kin}}/10,\,m_t^{\mathrm{kin}}/3]$ the variation of $\Gamma_t^{\mathrm{kin}}(\mu_{c})$ at N3LO reaches the level of one per-mile when $\mu$ is about $\mtkin/2$.
The number in the bracket right after the total value in \eqref{eq:explicitDWKin} corresponds to the change of the last two digits when $\mu$ and $\mu_{c}$ are varied as just described.

Lastly, we have also applied the recently proposed trace-anomaly-subtracted $\sigma$-mass definition~\cite{Chen:2025iul,Chen:2025zfa} for heavy quarks, which is not only scheme- and scale-invariant but also free from the leading IR-renormalon ambiguity.
With $m_t^{\sigma} = 159.0$~GeV obtained from $m_t = 172.69$~GeV using the three-loop conversion relation~\cite{Chen:2025iul}, we obtained the following result for the decay width of $t$-quark in $\sigma$-mass scheme: 
\begin{equation}\label{eq:explicitDWsiM}
\Gamma^{\sigma}_t(\mu=m_t^{\sigma}/2) = 
1.10470\, +\, 0.20790 \,+\, 0.01341 \,-\,0.00401 \,=\, 1.3220(30)~\text{GeV} 
\end{equation}
The number in the bracket right after the total value in \eqref{eq:explicitDWKin} quantifies the change of the last two digits when $\mu$ is varied from $m_t^{\sigma}/4$ to $m_t^{\sigma}$. 
Here there does not seem to be the accidental features regarding the behavior of the third-order correction and scale variation, in contrast to the result in the 5-flavor $\MSbar$ mass scheme mentioned before.

After restoring $V_{tb}$ and collecting the additional sources of systematic errors, our final result for the semi-leptonic decay width of $t$-quark reads\footnote{As noted in the end of the preceding subsection~\ref{sec:Tdecay_OS}, for the inclusive $\Gamt$, the finite $m_b$ and off-shell $W$ effects are largely canceled against the NLO electroweak correction to the point negligible in view of the level of precision in~\eqref{eq:GammaTfin}, %, at the sub-per-mille level, 
when they are taken into account in the multiplicative manner on top of the perturbative QCD corrections with massless $b$-quark and on-shell W-boson. They are thus not included in~\eqref{eq:GammaTfin}.}  
\begin{equation}\label{eq:GammaTfin}
\Gamma_t = 1.321(3)(2) \times |V_{tb}|^2 \,+\, 0.027\,(m_t - 172.69)\,~\text{GeV} 
\end{equation}
where the central value in the first term is obtained by averaging the stable results for $\Gamt$ in~\eqref{eq:explicitDWMS}, ~\eqref{eq:explicitDWKin} and \eqref{eq:explicitDWsiM}. 
The first and second numbers in brackets in the first term quantities, respectively, the error associated with the QCD ($\alpha_s$-renormalization) scale uncertainty\footnote{The number $(3)$ is obtained via taking the largest conventional scale uncertainties and the spread among the individual results.}  
and the error induced by the experimental uncertainty of the input QCD coupling $\alpha_s(m_Z)=0.1179\pm 0.0009\,$\cite{ParticleDataGroup:2022pth}. 
The second term parameterizes the main source of the error on $\Gamt$ originated from the experimental uncertainty of the input $t$-quark mass value.

\subsection{$\Gamma(\BXulv)$}
\label{sec:Bdecay}

The inclusive semi-leptonic decay width $\Gamma(\BXulv)$ provides a direct path to determining the CKM matrix element $|V_{ub}|$, crucial for our understanding of the Flavor Physics. 
This approach has the added merit of permitting a reliable theoretical treatment based on OPE,  where the non-perturbative corrections are suppressed as $1/m_b^2$ and can be expressed in terms of universal parameters of the heavy quark theory which are known with relatively small uncertainties (see, e.g.~refs.\cite{Uraltsev:1999rr,Hoang:1998hm,Bordone:2021oof,Gambino:2020jvv,Fael:2024rys}).
Consequently, one of the remaining major theoretical challenges for achieving a more precise extraction of $|V_{ub}|$ from $\Gamma(\BXulv)$ is the better theoretical understanding of the perturbative QCD corrections and the heavy quark masses. %The leading power dependence of the $b$-quark decay width on its mass is 5.
In this subsection we will provide our best theoretical prediction for the perturbative corrections to this quantity.

The application of the OPE method to the inclusive charmless semi-leptonic B-meson decays $\BXulv$ yields the following well-known formula~\cite{Bigi:1992su,Bigi:1993fe,Uraltsev:1999rr},
\begin{equation} \label{eq:gammaB2Xulv}
\Gamma\big(\BXulv\big)=\Gamma_0(m_b)\, \Big[
\mathbf{C}_{\mathrm{p}}\,\Big(
1 - \frac{\mu_\pi^2 - \mu_G^2}{2\, m_b^2}
\Big)
- 2\, \frac{\mu_G^2}{m_b^2} 
+ \mathcal{O}\big(\Lambda_{\mathrm{QCD}}^3/m_b^3\big)\Big]\,,
\end{equation}
where $\mathbf{C}_{\mathrm{p}}$ denotes the perturbative QCD correction factor, and $\mu_{\pi}^2$ and $\mu_{G}^2$ in the $1/m_b^2$-suppressed terms are, respectively, the expectation values of the kinetic and chromomagnetic operators in OPE or HQET (which encode the non-perturbative effects).
The perturbative QCD effects on $\Gamma\big(\BXulv\big)$ can be determined for a free $b$-quark as a power series in $\alpha_s$:%
\begin{eqnarray}
\label{eq:parametricDW}
\GamBulv &\equiv& \Gamma_0(m_b)\,\mathbf{C}_{\mathrm{p}}\,
= \Gamma_0(m_b)\, \Big[1 + \frac{\als}{\pi} \mathbf{c}_1 + \Big(\frac{\als}{\pi}\Big)^2 \mathbf{c}_2 + \Big(\frac{\als}{\pi}\Big)^3  \mathbf{c}_3 
+ \mathcal{O}(\als^4) \Big]\,, 
\end{eqnarray}%
where a prefactor $\Gamma_0 \equiv \frac{G^2_F\,  |V_{ub}|^2\, m_b^5\, A_{\mathrm{ew}}}{192 \pi^3}$ is introduced in the first line defined with the $b$-quark on-shell mass $m_b$, the Fermi constant $G_F = 1.166379\times 10^{-5}\,$GeV$^{-2}$, and the CKM matrix element $V_{ub} = 3.82 \times 10^{-3}$. 
$A_{\mathrm{ew}} = 1.01435$ denotes the currently known electroweak K-factor~\cite{Sirlin:1981ie,Fael:2024fkt}.

In terms of the quantities introduced in~\eqref{eq:GammaTsl}, the inclusive semi-leptonic decay width of a free $b$-quark, $\GamBulv$ can be specified as 
\begin{eqnarray} \label{eq:Gamma_b2ulv}
\GamBulv 
&=& 
\int^{m_b^2}_{0} \frac{\mathrm{d} m_w^2}{2 \pi} 
\frac{2\, m_W \, \Gamma_{W \rightarrow \ell \bar{\nu}_{\ell}} }{ 
\big(m_w^2 - m_W^2 \big)^2 + \big(m_W \Gamma_W^{tot} \big)^2} 
\, \tilde{\Gamma}_{b}(m^2_b, m_w^2) \, \Big|_{m_b \ll m_W}  \nonumber\\
&=& \frac{G_F }{3 \sqrt{2} \pi } \,
\int^{m_b^2}_{0} \frac{\mathrm{d} m_w^2}{2 \pi} \, 
\tilde{\Gamma}_{b}(m^2_b, m_w^2)  \nonumber\\
 &=&  \frac{G_F^2 m_b^5}{192 \pi^3}  \,+\, \mathcal{O}(\als)\, \,,
\end{eqnarray}
where we have omitted the CKM matrix elements and taken the leading-order approximation in electroweak interaction. 
This is similar to~\eqref{eq:GammaTsl} for $t$-quark except for now in the limit of a heavy W-boson.

In our calculation of the QCD corrections to $\GamBulv$, the diagrams with closed $b$-quark loops are not included, and all $n_l=4$ light flavors are taken massless. 
The contributions to $\GamBulv$ from the diagrams with $b$-quark loops start from two-loop order, and have been determined in refs.~\cite{vanRitbergen:1999gs,Fael:2023tcv}, which account for a few $10^{-4}$ of the total N3LO result. 
To be complete, these small contributions are, nevertheless, included in our final results for $\Gamma\big(\BXulv\big)$ to be presented below. 
(In addition, the leading-color part was derived analytically up to N3LO result in ref.~\cite{Chen:2023dsi}.)  
The perturbative coefficients $\mathbf{c}_n$ in \eqref{eq:parametricDW} depend on renormalization scale $\mu$ in general, and also logarithmically on the $b$-quark mass, due to the $\MSbar$ renormalization of the QCD coupling $\als$ (as well as the $b$-quark mass).
We begin with a documentation of the values of the $b$-quark masses and $\als$ used.
With inputs for the 5-flavor $\alpha_s(m_Z) = 0.1179$ at the Z-pole mass $m_Z = 91.1876$ GeV and the PDG average value $\mbms(\mbms) = 4.18^{+0.03}_{-0.02} \mathrm{\,GeV}$ (in 5-flavor scheme), 
we obtain $\als(\mu=\mbms(\mbms)) = 0.22424$ using the $\als$-running at four-loop order.\footnote{The values for the b-quark on-shell mass $m_b$ are determined to be $4.78\,$, $4.92\,$ and $5.06\,$GeV using the 2-, 3- and 4-loop mass conversion formula from the input $\mbms(\mbms) = 4.18\,$GeV.}
Numbers for the 5-flavor $\als(\mu)$ and $\mbms(\mu)$ at other scales can be readily obtained by solving the renormalization-group equations.

We obtain the following numerical results for the perturbative coefficients defined in the on-shell scheme at the scale $\mu=m_b$:
\begin{eqnarray}
\label{eq:Cos_nl}
\mathbf{c}_1 = -2.41307\,, \,
\mathbf{c}_2 = -29.8588 + 2.1469\, n_l \,, \,
\mathbf{c}_3 = -550.014 + 79.0445 \, n_l - 2.30649\, n_l^2\,,\nonumber\\
\end{eqnarray}%
where $n_l=4$ light flavors are taken massless and the $b$-quark loops are not included. % $\mathbf{c}_0 = 1$ 
Including the contributions to $\GamBulv$ from the diagrams with $b$-quark loops~\cite{vanRitbergen:1999gs,Fael:2023tcv}, the complete results read
\begin{eqnarray}
\label{eq:Cos_full}
\mathbf{c}_1 = -2.41307\,, \,
\mathbf{c}_2 = -29.8831 + 2.1469\, n_l \,, \,
\mathbf{c}_3 = -549.046 + 79.0402 \, n_l - 2.30649\, n_l^2\,,\nonumber\\
\end{eqnarray}%
which differ from \eqref{eq:Cos_nl} only at per-mile level. 
Insert $n_l=4$ into \eqref{eq:Cos_full} and we obtain 
\begin{eqnarray}
\label{eq:Cos}
\mathbf{c}_1 = -2.41307\,, \quad \mathbf{c}_2 = -21.2955 \,, \quad \mathbf{c}_3 = -269.788\,,
\end{eqnarray}%
which are used in the following discussions of our final results for $\Gamma\big(\BXulv\big)$.

It is known in literature that the perturbative coefficients $\mathbf{c}_n$ for $\GamBulv$ expressed in term of on-shell renormalized $b$-quark mass receive contributions that grow rapidly ($\sim n!\,r^n$ power-factorial-like behavior) at higher orders corresponding to a singularity related to the infrared renormalon issue of the on-shell (pole) mass definition~\cite{Beneke:1994qe,Smith:1994id,Ball:1995wa,Smith:1996xz,Bigi:1996si,Beneke:1998ui,Beneke:2016cbu,Hoang:2017btd,FerrarioRavasio:2018ubr}. 
We have isolated the leading $\alpha_s^n\,\beta_0^{n-1}$-enhanced terms for $n=2,3$ in the perturbative corrections to $\GamBulv$  in the on-shell scheme, in the large-$n_f$ approximation~\cite{Beneke:1994qe,Broadhurst:1994se}. 
After normalized w.r.t the perturbative coefficient of the first $\alpha_s$-order term, the coefficients of $\alpha_s^2\,\beta_0$ and $\alpha_s^3\,\beta_0^2$ agree with the predictions given in ref.\cite{Ball:1995wa}. 
The above issue of rapid growth of the high-order perturbative corrections to $\GamBulv$ is known~\cite{Beneke:1994qe,Bigi:1996si,Smith:1994id,Ball:1995wa,Smith:1996xz,Hoang:1998hm,vanRitbergen:1999gs} to be tempered by rewriting $\GamBulv$ in terms of appropriate short-distance or threshold masses.

\begin{figure}[htbp]
\begin{center}
\includegraphics[scale=0.9]{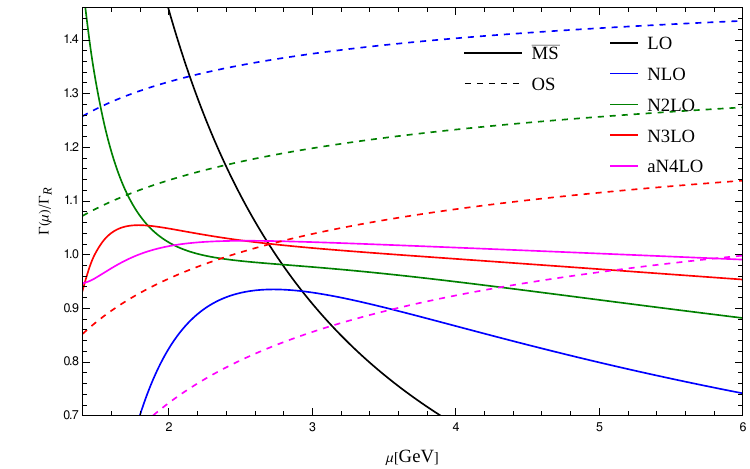}
\caption{The scale dependence of the fixed-order results for $\GamBulv$ and $\GamBulvMS$ in $\mu \in [1.4, 6]\,$GeV. 
The numbers shown in the y-axis for the decay width are conveniently normalized by a chosen reference value $\Gamma_R \equiv 6.59 \times 10^{-16}$~GeV.
}
\label{fig:TotWidthMuD_b2ulv_MSOS}
\end{center}
\end{figure}

Since the convergence behavior of the above perturbative series in the on-shell scheme is not good (also implied by its relatively large scale uncertainty, see below), we re-express the results \eqref{eq:Cos} for $\GamBulv$ in terms of the $\MSbar$ $b$-quark mass. 
At the scale $\mu=\mbms(\mbms)$, we obtain  
\begin{eqnarray}
\label{eq:Cms}
\bar{\mathbf{c}}_0 = 1\,,  \quad \bar{\mathbf{c}}_1 = 4.25360\,, \quad \bar{\mathbf{c}}_2 = 26.7848 \,, \quad\bar{\mathbf{c}}_3 = 188.937\,.
\end{eqnarray}%
As noted in refs.~\cite{Bigi:1996si,vanRitbergen:1999gs}, the first few perturbative corrections in the $\MSbar$ scheme are tamed to be roughly a geometric-power-like series.
For the sake of reference, our numerical result for $\GamBulv$ in 5-flavor $\MSbar$ scheme at $\mu=\mbms(\mbms)$ reads:
\begin{eqnarray}
\label{eq:Gms}
\GamBulvMS &=& \Gamma_0(\mbms) \, 
\big(
1 \,+\, 0.3036 \,+\, 0.1367 \,+\, 0.0687 
%\,+\, {\color{blue}0.03433}
\big)
\end{eqnarray}%
where $\Gamma_0(\mbms = 4.18) = 4.3164 \times 10^{-16}$~GeV.
The regularity observed in \eqref{eq:Gms} is that every one more perturbative order higher in this $\MSbar$ series, the size of the term is reduced roughly by a relative factor $1/2$.
In fig.~\ref{fig:TotWidthMuD_b2ulv_MSOS}, we illustrate the residual scale dependence of our fixed-order results for $\GamBulvMS$ in $\mu \in [1.4, 6]\,$GeV, which are exact up to N3LO, both in the $\MSbar$ and on-shell mass schemes for comparison.
(We have used the renormalization group equations at four-loop order for the running of both $\alpha_s$ and $\MSbar$-mass $\overline{m}_t$ in the 5-flavor scheme in these plots.)
It is obvious that both the convergence and the residual scale dependence of the perturbative series of $\GamBulvMS$ are improved compared to that of $\GamBulv$. 
Moreover, the size of the scale dependence of $\GamBulvMS$ is gradually reduced once higher-order  perturbative corrections are included.
In particular, the scale uncertainty for $\GamBulvMS$ corresponding to $\mu \in [\mbms/2, 2\,\mbms]$ up to $\mathcal{O}(\als^3)$ is about $[+6\%, -8\%]$. 
We note that the scale uncertainty may be further reduced to $[+0.8\%, -5\%]$ upon incorporating an  approximated $\als^4$-order correction estimated by exploiting the aforementioned geometric-series-like pattern (which is indicated in~fig.~\ref{fig:TotWidthMuD_b2ulv_MSOS} by the solid magenta line). 
It is known that the scale $\mu \sim m_b$ is not the best or physically the most reasonable choice for $\GamBulvMS$, since a typical energy release into the hadronic decay product is much less than $m_b$.
From this viewpoint, scales below $m_b$ shall be more natural for $\GamBulvMS$.
In this regard, we notice that the $\MSbar$ results at different perturbative orders in fig.~\ref{fig:TotWidthMuD_b2ulv_MSOS} tend to become close to each other for $\mu \in [2\,,\,3]$~GeV; 
in particular, around $\mu \sim m_b/2 =  2.5\,$ GeV, the estimated $\mathcal{O}(\als^4)$ correction to $\GamBulvMS$ is almost vanishing.
~\\

Despite the considerable improvement in the behavior of the perturbative series for $\GamBulvMS$ in $\MSbar$ mass scheme, as shown in fig.~\ref{fig:TotWidthMuD_b2ulv_MSOS}, its conventional scale uncertainty at N3LO is still as large as $[+6\%, -8\%]$.
We thus seek for further improvements by rewriting $\GamBulv$ in terms of the kinetic mass~\cite{Bigi:1994ga,Bigi:1996si,Czarnecki:1997sz} $\mbkin$ for $b$-quark.
Employing the three-loop relation between the pole mass and kinetic mass \cite{Fael:2020iea}, and the input value $\mbkin[1] \equiv \mbkin(\mu_c=1\,\mathrm{GeV}) = 4.554 \pm 0.018$~GeV from the HFLAV collaboration~\cite{HFLAV:2019otj}, 
we obtain 
\begin{eqnarray}
\label{eq:Gkin}
\GamBulvK &=& \Gamma_0(\mbkin[1]) \, 
\Big(
1\, -\, 0.0271 \,+\, 0.0251 \,+\, 0.0218
\Big)
\nonumber\\
&=&  6.7565 \times 10^{-16}\,\text{GeV}\,.
\end{eqnarray}%
at the wittily chosen $\mu=\mbkin[1]/2$ for the 4-flavor $\alpha_s(\mu)$ in view of the discussions in ref.~\cite{Bordone:2021oof}.
We have cross-checked with ref.~\cite{Uraltsev:1999rr} on the full $\mathcal{O}(\als)$ and the partial or approximated $\mathcal{O}(\als^2)$ results. 
Our complete $\mathcal{O}(\als^2)$ and $\mathcal{O}(\als^3)$ results in the kinetic mass scheme are new. 
In \eqref{eq:Gkin}, the purely third-order correction increases the N2LO result by about $2\%$.

\begin{figure}[htbp]
\begin{center}
\includegraphics[scale=0.9]{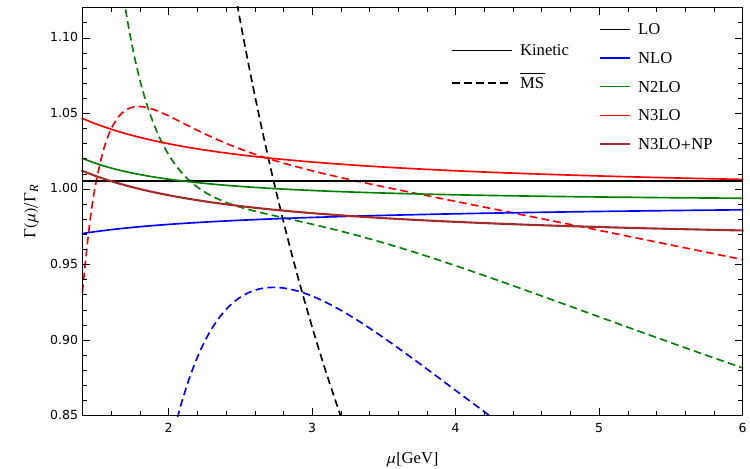}
\caption{The scale dependence of the fixed-order results for $\GamBulvK$ and $\GamBulvMS$ in $\mu \in [1.4, 6]\,$GeV. 
The numbers shown in the y-axis for the decay width are conveniently normalized by a chosen reference value $\Gamma_R \equiv 6.59 \times 10^{-16}$~GeV.}
\label{fig:TotWidthMuD_b2ulv_MSKin}
\end{center}
\end{figure}

In fig.~\ref{fig:TotWidthMuD_b2ulv_MSKin}, we illustrate the residual scale dependence of our fixed-order results for $\GamBulv$ reformulated using the kinetic mass $\mbkin[1]$ in $\mu \in [1.4, 6]\,$GeV, where the results in $\MSbar$ scheme are included for comparison.\footnote{Note that the $\alpha_s$ in the $\MSbar$ results is defined in the 5-flavor scheme as in fig.~\ref{fig:TotWidthMuD_b2ulv_MSOS}, while the $\alpha_s$ in the results with kinetic mass is defined in the 4-flavor scheme; 
for both cases the four-loop running formula is used in fig.~\ref{fig:TotWidthMuD_b2ulv_MSKin}.}
The improvement in the behavior of the perturbative series of $\GamBulvK$ over that of $\GamBulvMS$ is clearly visible in fig.~\ref{fig:TotWidthMuD_b2ulv_MSKin}.
In particular, the conventional scale uncertainty of $\GamBulvK$ at N3LO determined via varying $\mu$ between $\mbkin[1]/2$ and $2\,\mbkin[1]$ is reduced to the level of 2\%, which is more than a factor-of-6 reduction compared to the $\MSbar$ counterpart.
However, it might be a bit confusing at first sight to notice that, though quite small, the conventional scale bands do not decrease further as one goes from N2LO to N3LO. 
The discussions on the behaviors of the perturbative QCD corrections to the so-called $q^2$-spectrum in subsection~\ref{sec:lpmass_b2ulv} will offer an illuminating insight into this issue.

It is now straightforward to incorporate the $1/m_b^2$-power suppressed non-perturbative corrections, parameterized in terms of two universal parameters in \eqref{eq:gammaB2Xulv} well-defined in OPE or HQET. 
Inserting our three-loop perturbative result \eqref{eq:Gkin} and the fit values for the non-perturbative parameters $\mu_{\pi}^2 = 0.477 \pm 0.056 \mathrm{\,GeV^2}$~and $\mu_{G}^2 = 0.306 \pm 0.050 \mathrm{\,GeV^2}$~\cite{Bordone:2021oof}
into the master formula \eqref{eq:gammaB2Xulv}, we finally obtain the state-of-the-art theoretical prediction for the charmless semi-leptonic decay width of the B-meson:
\begin{equation}
\label{eq:GB2Xulv}
\Gamma(\BXulv) =  \frac{|V_{ub}|^2}{|3.82\times 10^{-3}|^2}\,\big( 6.53 \,\pm 0.12 \, \pm 0.13\, \pm 0.03\, \big) \times 10^{-16}\,\text{GeV}\,,   
\end{equation}
%(the number 6.53 was determined using $|V_{ub}|^2 = |3.82\times 10^{-3}|^2$)
where the first error corresponds to the fit error of the input kinetic mass $\mbkin[1]$; the second error is assigned from the residual scale uncertainty estimated conservatively by the variation of \eqref{eq:GB2Xulv} for $\mu$ between $\mbkin[1]/2$ and $2\,\mbkin[1]$;
the third error is related to the errors on the non-perturbative fit parameters.\footnote{Additional errors of other sources are much smaller compared to the total of the explicitly listed ones.
For example, assuming a $3\%$ error for the value of $\alpha_s(\mbkin[1]/2)$, estimated using the 
input $\alpha_s(m_z) = 0.1179\pm 0.0009$~\cite{ParticleDataGroup:2022pth} with four-loop running formula, leads to a change of $\Gamma(\BXulv)$ less than two per-miles. %$6.53 \pm 0.017$
}
Adding these errors in quadrature leads to about $3\%$ error on our final number for $\Gamma(\BXulv)$ in \eqref{eq:GB2Xulv}. 
The residual scale dependence of our result for $\Gamma(\BXulv)$ is plotted by the brown dot-dashed line in fig.~\ref{fig:TotWidthMuD_b2ulv_MSKin}.
We note that the purely third-order correction determined in this work increases the N2LO result for by $\Gamma(\BXulv)$ about $2\%$.
It is interesting to note that the central value of our theoretical result \eqref{eq:GB2Xulv} lies very close to the averaged central values of the experimental measurements for $\Gamma(\BXulv)$ in \eqref{eq:GBexp}, differing from each other by less than $1\%$.
For the sake of reference, the averaged value of semi-leptonic B$^{\pm/0}$-meson decay width reads 
\begin{equation}
\label{eq:GBexp}
\Gamma_{\mathrm{exp}}(B \rightarrow X_u \ell \bar{\nu}_{\ell}) = 6.59 \times 10^{-16}\,\mathrm{GeV}\,.    
\end{equation} 
To be more specific, we derived the above value by combining the following data. 
With the lifetime $(1638 \pm 4) \times 10^{-15}\,s$ for the $B^{\pm}$-meson and the branch ratio $(1.65 \pm 0.21) \times 10^{-3}$ for the $X_u \ell \bar{\nu}_{\ell}$~\cite{Belle:2021eni}, one obtains the semi-leptonic B-meson decay width $\Gamma_{\mathrm{exp}}(B^{\pm} \rightarrow X_u \ell \bar{\nu}_{\ell}) = 6.63 \times 10^{-16}\,$GeV. 
With the lifetime $(1517\pm 4) \times 10^{-15}\,s$ for the $B^{0}$-meson and the branch ratio $(1.51 \pm 0.19)  \times 10^{-3}$ for the $X_u \ell \bar{\nu}_{\ell}$, one obtains the semi-leptonic B-meson decay width $\Gamma_{\mathrm{exp}}(B^{0} \rightarrow X_u \ell \bar{\nu}_{\ell}) = 6.55 \times 10^{-16}\,$GeV.

\subsection{Lepton-pair invariant mass spectrum}
\label{sec:lpmass}

The lepton-pair invariant-mass or $q^2$-spectrum in $\BXulv$ was proposed~\cite{Bauer:2000xf,Bauer:2001rc} as one of the precision observables for extracting the CKM matrix element $|V_{ub}|$, and has been measured at Belle II~\cite{Belle:2003vfz,Belle:2021ymg}.
To this end, certain experimental cuts shall be imposed on the invariant mass squared $q^2$ of the lepton pair, such as $q^2 > (m_B - m_D)^2$, to suppress the $\BXclv$ background while still keeping a significant fraction of the $\BXulv$ signal events. 
Ref.~\cite{Bauer:2000xf} showed that measuring this observable with appropriate cuts may yield a precision determination of $|V_{ub}|$ %"model" here refers to the model for the non-perturbative shape function involved in the lepton-energy spectrum especially close to the end-point 
with suppressed dependence on the shape-function models and allowing more signal events, as compared to the approaches based on severe cuts on lepton-energy and hadronic invariant mass spectra. %to suppress the effect of the not-well known non-perturbative shape functions. 
It was further shown~\cite{Bauer:2001rc} that applying appropriate combined cuts on the invariant masses of both the leptonic and hadronic system is more advantageous in order to keep the theoretical uncertainties under control while retaining a larger data sample of the $\BXulv$ events. 
Measurements of $|V_{ub}|$ following this approach had been performed at Belle and Belle II~\cite{Belle:2003vfz,HFLAV:2016hnz}.
Although the phase-space regions with large $q^2$ are the most precise and sensitive fiducial region for the inclusive measurement of $\BXulv$, they are unfortunately also the regions where both the non-perturbative effects and perturbative QCD corrections are kinematically enhanced. 
The perturbative uncertainties constitute an important source of errors in this approach\cite{Belle:2003vfz}, albeit not listed as the largest in the past  analysis~\cite{HFLAV:2016hnz}. 
However, regarding this aspect, our following discussions on the high-order perturbative corrections to the $q^2$-spectrum in $\BXulv$ (in different quark mass schemes) may be useful for gaining some new insights into the relatively large differences in the inclusive determination of $|V_{ub}|$ extracted using this approach as compared to using the others~\cite{HFLAV:2016hnz,Belle-II:2018jsg}.

\begin{figure}[htbp]
\begin{center}
\includegraphics[scale=0.75]{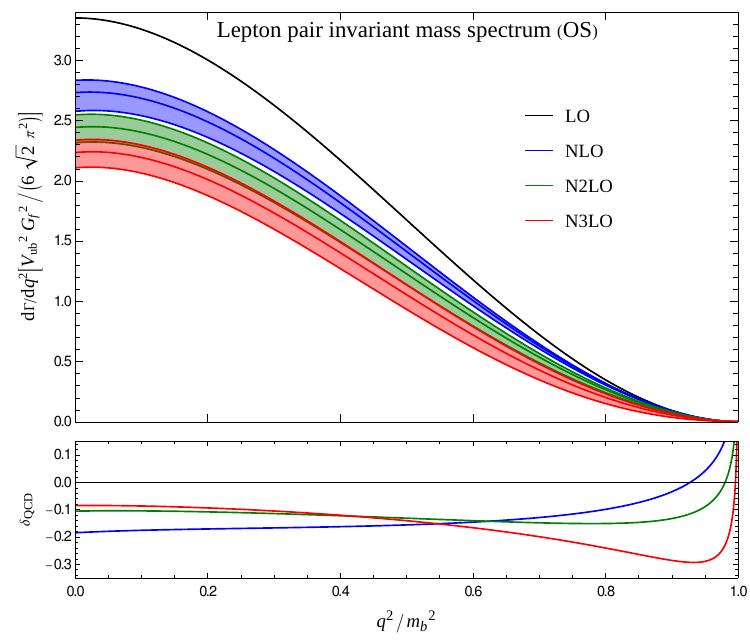}
\caption{The perturbative results for $q^2$-spectrum in $\bulv$ up to $\mathcal{O}(\alpha_s^3)$ determined in the $b$-quark pole mass scheme with the conventional scale-variation bands obtained by varying $\mu$ from $\mbos/2$ to $\mbos$. The horizontal x-axis is the rescaled $q^2$ normalized by $\mbos^2$, and a constant factor is pulled out as the unit for the values shown on the y-axis.}
\label{fig:b2ulv_dGamdq2_OS}
\end{center}
\end{figure}

In this section, we discuss the perturbative results for the lepton-pair invariant mass spectrum $\frac{\mathrm{d} \GamBulv}{\mathrm{d} q^2}$ of the parton-level decay process $\bulv$, referred to below simply as $q^2$-spectrum for short, determined in different $b$-quark mass schemes.
We start with the results determined using the $b$-quark pole mass $\mbos$. 
To be definite, in term of the inclusive decay width~\eqref{eq:Gamma_b2ulv}, $\frac{\mathrm{d} \Gamma_{b \rightarrow u}}{\mathrm{d} q^2}$ to be presented can be defined by 
\begin{eqnarray}\label{eq:Gamma_b2ulv_O_q2}
\GamBulv 
&=&\int^{\mbos^2}_{0} \frac{\mathrm{d} \GamBulv }{\mathrm{d}\, q^2} \, \mathrm{d}\, q^2\, \nonumber \\
&=& \int^{\mbos^2}_{0} \big( f_0 \,+\, \als\, f_1 \,+\, \als^2\, f_2 \,+\, \als^3\, f_3  \big) \, \mathrm{d}\, q^2\,,
\end{eqnarray}
where the shorthand notations $f_i$ are introduced for the perturbative correction coefficients at the differential level.
As in the previous subsection~\ref{sec:Bdecay} on $\bulv$, we work with a 5-flavor QCD Lagrangian where only the $b$-quark is kept massive. In particular, the virtual $c$-quarks involved in the contributing diagrams are approximated massless.\footnote{With the explicit results for $\BXclv$~\cite{Dowling:2008mc,Fael:2020tow} the mass effect of the $c$-quark loops changes the $(\alpha_s/\pi)^2$-coefficient by about $4.5\%$ in the on-shell scheme, and its impact on the total decay width is below $0.3$-permile. We expect the similar in $\BXulv$, as the $c$-quark loop enters from this order only via the gluon-self-energy correction.}  
With the same input parameters, which include the 5-flavor $\alpha_s(m_Z)=0.1179\pm 0.0009\,$\cite{ParticleDataGroup:2022pth} at the Z-pole mass $m_Z = 91.1876$ GeV and the PDG average value $\mbms(\mbms) = 4.18^{+0.03}_{-0.02} \mathrm{\,GeV}$ (in 5-flavor scheme), the $b$-quark on-shell or pole mass $\mbos$ is determined to be $4.78\,$ and $4.92\,$GeV, respectively, using the 2- and 3-loop mass conversion formula~\cite{Melnikov:2000qh}.
Our numerical results for $\frac{\mathrm{d} \Gamma_{b \rightarrow u}}{\mathrm{d} q^2}$ with the $b$-quark pole mass $\mbos = 4.92\,$GeV are plotted in fig.~\ref{fig:b2ulv_dGamdq2_OS}, up to $\mathcal{O}(\alpha_s^3)$ together with the conventional QCD-scale variation bands indicated in different colors.
The $q^2$-spectrum is cross-checked against the expanded result up to $\mathcal{O}(\alpha_s^2)$~\cite{Czarnecki:2001cz} valid in the large-$q^2$ region.  
The leading-color part of the integrated results in the on-shell scheme has been further cross-checked with those reported in ref.~\cite{Chen:2023dsi}.
The integrated form of the fermionic-loop dependent part has also been cross-checked with the ref.~\cite{Fael:2023tcv}.

As shown by the upper panel of fig.~\ref{fig:b2ulv_dGamdq2_OS}, the $q^2$-spectrum at various orders and scales considered all turn to vanish at the boundary where $q^2$ takes its maximum $\mbos^2$   
and hence the total energy of the hadronic system virtually vanishes. 
In the lower panel for $\delta_{\mathrm{QCD}}$, the ratio of the pure $\alpha_s^{N}$-th correction to the total perturbative result up to the previous $\alpha_s^{N-1}$-order is plotted for $N=1,\,2,\,3$.  
Namely $\delta_{\mathrm{QCD}}^{(N)} = \frac{\alpha_s^N\, f_N}{\sum_{i=0}^{N} \alpha_s^i\, f_i}$ with $f_i$ defined in \eqref{eq:Gamma_b2ulv_O_q2} (We simply take the symbol $\delta_{\mathrm{QCD}}$ omitting the superscript as a collective notation for these rates).
It shows that $\delta_{\mathrm{QCD}}$ determined in the pole mass scheme is negative in most of the $q^2$-region at each perturbative order, which is expected based on the perturbative QCD effect on $\GamBulv$ discussed in section~\ref{sec:Bdecay}.
These perturbative QCD corrections are quite sizable, 
and the result at $\mathcal{O}(\alpha_s^{N})$ at the central scale $\mu=\mbos$ lies outside the conventional scale-variation band of the previous order (i.e.~those obtained by varying $\mu$ from $\mbos/2$ to $\mbos$). 
Moreover, there is almost no overlap between the shown conventional scale-variation bands.
Lastly we see that in the on-shell scheme $\delta_{\mathrm{QCD}}$ varies rapidly as $q^2$ approaches its maximum $\mbos^2$, more so at higher perturbative orders, even though the absolute values of the distribution become small in this region.

Based on the detailed discussion of the perturbative QCD corrections to $\GamBulv$ in section~\ref{sec:Bdecay}, the pole-mass scheme is clearly not the best choice for describing the (differential) decay width for $\bulv$.
However, it turns out that even limited to just perturbative corrections, changing quark-mass schemes, consistently and thoroughly is much less straightforward but entails additional subtlety for a full differential spectrum compared to the previous cases of inclusive decay widths as well as their inclusive moments. 
The following subsection will be devoted to a detailed exposition of the technical subtitles involved, as we have not yet seen any discussion on these points for $q^2$-spectrum elsewhere.

\subsubsection{Boundary terms in the perturbative re-expansion of an integral}
\label{sec:lpmass_reformula}

The task of reformulating the $q^2$-spectrum from the original form determined in the on-shell scheme into a new perturbative form using an alternative quark-mass definition, where there is no more \textit{any} explicit reference to the pole mass, is conceptually similar to the cases of inclusive quantities (with much simpler mass dependence) but entails new technical subtleties. %when insisted at the fully differential level covering the whole region
The treatment involved can be abstracted into the following mathematical question. 
Let us consider a prototype parametric integral of a finite function of the mass $m$,  
\begin{equation}\label{eq:integral}
F(\alpha_s\,,\, m) = \int_{0}^{m}\, f(\alpha_s\,,\, m\,;\, x) \, \mathrm{d} x 
\end{equation}
where the integration over $x$ is regular in the integration domain $(0, m)$.  
The lower bound of the integration does not need to be 0, but can take any $m$-independent value within the domain where the known $f(\alpha_s\,,\, m\,;\, x)$ remains valid. 
The dependence of $F(\alpha_s\,,\, m)$ on $\alpha_s$ is truncated to a fixed power order $\mathcal{O}(\alpha_s^N)$ of interest, e.g.~$N=3$ in present applications. 
Furthermore, we are provided with the following (finite) perturbative relation between the original mass $m$ and a new one $0 < m_0 < m$: 
\begin{equation}\label{eq:transformation}
m = C(\alpha_s,\, m_0) = m_0 \,+\, \sum_{n=1}^{N} c_n\, \alpha_s^n \,+\, \mathcal{O}(\alpha_s^{N+1}) \equiv m_0 \,+\, \delta m 
\end{equation}
where the perturbative coefficients $c_n$ are functions of $m_0$ in general, and $\delta m$ admits a perturbative expansion starting from $\mathcal{O}(\alpha_s)$.
The task is to perturbatively re-expand the mass dependence of $F\big(\alpha_s\,,\, m_0 \,+\, \delta m \big)$ in terms of $m_0$ and truncate consistently to the same perturbative order $N$ of interest in $\alpha_s$.  
For the practical applications in question, $m$ shall be the $b$-quark pole mass $\mbos$ while $m_0$ may be its alternative masses such as $\mbms$ or $\mbkin$.
For the convenience of later reference, we may denote the resulting function by $\tilde{F}(\alpha_s, m_0)$ which shall exactly equal $F\big(\alpha_s\,,\, m_0 \,+\, \delta m \big)$ %in value 
\textit{if} no perturbative expansion and truncation is ever performed (or resummed to all perturbative orders assuming convergence of the perturbative series).
It is straightforward to have this done for $F\big(\alpha_s\,,\, m \big)$, where its partial derivatives $\frac{\partial^n\, F}{\partial\, m^n}$ up to $N$-th order will be involved, provided its partial dependence on $m$ around the new expansion-point $m_0$ is regular and given explicitly (namely admits a convergent Taylor-series expansion), just like done in subsection~\ref{sec:Tdecay} and~\ref{sec:Bdecay}.
This convergence property is assumed below, at least up to the perturbative order $N$ in question.

Since both the $m$-dependence of the integral around $m_0$ and the integration over $x$ are regular, the aforementioned perturbative re-expansion may be performed at the integrand level before the integration over $x$. 
Consequently, this will lead to an integrand or distribution whose integration gives back 
\begin{equation}\label{eq:integralexp}
\tilde{F}(\alpha_s, m_0) = \int_{0}^{m_0}\, \tilde{f}(\alpha_s\,,\, m_0\,;\, x) \, \mathrm{d} x\,,
\end{equation}
which may differ from $F(\alpha_s\,,\, m)$ but only due to perturbative truncation to certain fixed orders in $\alpha_s$. 
To this end, extra care is required because the $m$-dependence in the parametric integral representation~\eqref{eq:integral} appears not only in the integrand $f(\alpha_s\,,\, m\,,\, x)$, but also via the integration boundary!
Therefore, a thorough and consistent re-expansion of the $m$-dependence of the r.h.s.~of~\eqref{eq:integral} can be separated into the following two parts:
\begin{eqnarray}\label{eq:integralpowexp}
\int_{0}^{m}\, f(\alpha_s\,,\, m\,;\, x) \, \mathrm{d}\, x 
&=& \int_{0}^{m_0}\, f(\alpha_s\,,\, m_0 \,+\, \delta_m \,;\, x) \, \mathrm{d} x
\,+\, 
\int_{m_0}^{m}\, f(\alpha_s\,,\, m_0 \,+\, \delta_m \,;\, x) \, \mathrm{d} x  \nonumber\\ 
&=& \int_{0}^{m_0}\, 
\sum_{n=0}^{\infty} \frac{1}{n!}\, \frac{\partial^n\, f(\alpha_s\,,\, m \,;\, x)}{\partial\, m^n}\Big|_{m=m_0} \, \big(\delta m\big)^{n} \, \mathrm{d} x  
\,+\, 
\int_{m_0}^{m}\, f(\alpha_s\,,\, m_0 \,+\, \delta_m \,;\, x) \, \mathrm{d} x \, \, \nonumber\\
\end{eqnarray}
where $\delta m = m - m_0$ defined in \eqref{eq:transformation} starts from $\mathcal{O}(\alpha_s)$.
The first part, with $x$ integrated over $[0,\, m_0]$\footnote{The lower bound of the integration~\eqref{eq:integralexp} and~\eqref{eq:integralpowexp} needs not to be 0, but can take any $m$-independent value within the domain where the known $\tilde{f}(\alpha_s\,,\, m_0\,;\, x)$ remains valid, which practically means here that it is less than $m_0$.}, takes the normal form that is usually expected, where all re-expansion involves only the local $m$-dependence of the regular integrand $f$ at given $x$.

The second part $\int_{m_0}^{m}\, f(\alpha_s\,,\, m_0 \,+\, \delta_m \,;\, x) \, \mathrm{d} x$ is peculiar,  for it has an integration domain $\delta m = m - m_0$ of the order $\mathcal{O}(\alpha_s)$ as the small expansion parameter $ \alpha_s $ itself.
Given this feature and in view of the perturbative series expansion in $\alpha_s$ in the end, one may perform a Taylor expansion of $ f(\alpha_s\,,\, m_0 \,+\, \delta_m \,;\, x)$ in $x$ around $x = m_0$, and then complete the integration in $x \in [m_0, m]$ term-by-term. %each gives a polynomial factor in $\delta m$ 
To be explicit, 
\begin{eqnarray}\label{eq:boundaryterm}
\int_{m_0}^{m}\, f(\alpha_s\,,\, m_0 \,+\, \delta_m \,;\, x) \, \mathrm{d} x  
&=& \int_{m_0}^{m}\, 
\sum_{k=0}^{\infty} \frac{1}{k!}\, \frac{\partial^k\, f(\alpha_s\,,\, m_0 \,+\, \delta_m  \,;\, x)}{\partial\, x^k}\Big|_{x=m_0} \, \big(x - m_0\big)^k\, \mathrm{d} x \, \nonumber\\
&=& 
\sum_{k=0}^{\infty} \frac{1}{(k+1)!}\, \frac{\partial^k\, f(\alpha_s\,,\, m_0 \,+\, \delta_m  \,;\, x)}{\partial\, x^k}\Big|_{x=m_0} \,\, \big(\delta m \big)^{k+1}\,,
\end{eqnarray}
where each derivative $\frac{\partial^k\, f(\alpha_s\,,\, m_0 \,+\, \delta_m  \,;\, x)}{\partial\, x^k}\Big|_{x=m_0}$ may be further explicitly expanded in $m$ around $m_0$, up to the perturbative order in question. 
Consequently, what are eventually involved in the perturbative series~\eqref{eq:boundaryterm} are the value $f(\alpha_s\,,\, m_0\,;\, x=m_0)$ and its partial derivatives evaluated right at this point. 
Incorporating this class of additional terms originating from the $m$-dependence of the integration boundaries, or \textit{boundary-effect terms} for short, is necessary to \textit{exactly} maintain the original mathematical equality~\eqref{eq:integral} in the perturbatively re-expanded form. %unless they happen to be zero up to the perturbative order in question. 
In other words, they must be included if the integration of the resulting perturbatively re-expanded distribution $\tilde{f}(\alpha_s\,,\, m_0\,;\, x)$ shall give back the corresponding re-expanded $\tilde{F}(\alpha_s, m_0)$ --- which can be done using the given $F(\alpha_s\,,\, m)$ without making any reference to this integral representation --- namely, the same result obtained irrespective of whether the perturbative re-expansion in $\alpha_s$ is done before or after the integration in $x$. 
~\\

Let us be more specific about the boundary-effect terms for the $q^2$-spectrum in $b \rightarrow u \ell \bar{\nu}_\ell$ introduced in~\eqref{eq:Gamma_b2ulv_O_q2}. 
For brevity, we omit the arguments of the perturbative coefficients $f_i(q^2, \mbos)$, and use $m$ as a shorthand for $\mbos$.  
The explicit perturbatively expanded form of the boundary-effect terms defined in~\eqref{eq:boundaryterm} reads~\footnote{Note that the upper bound of $q^2$ in the integral~\eqref{eq:Gamma_b2ulv_O_q2} is $m^2$, and this point was properly taken into account in deriving the result~\eqref{eq:boundarytermRefine} following the idea illustrated using the prototype integral~\eqref{eq:integral}.} up to $\Oals{3}$:
\begin{align}\label{eq:boundarytermRefine}
& 2 \, \alpha_s\, c_1\, f_0\, m_0 +\alpha_s^2 \
\biggl(c_1^2\, \Bigl(2\, m_0^2\,
\frac{\partial f_0}{\partial q^2}+ 2\, m_0 \frac{\partial f_0}{\partial m}+f_0\Bigr)+2\,m_0 \left(c_2\, f_0+c_1\, f_1\right)\biggr) \nonumber \\
&+ \alpha_s^3 \biggl( c_1^3\, \Bigl(\frac{4}{3}\, m_0^3\, \frac{\partial^2 f_0}{\partial (q^2)^2}+2\, m_0^2\, \frac{\partial^2 f_0}{\partial q^2\, \partial m}+m_0\,
\frac{\partial^2 f_0}{\partial m^2}+2\, m_0\, \frac{\partial f_0}{\partial q^2}+\frac{\partial f_0}{\partial m}\Bigr) \nonumber \\ 
&\qquad +\,c_1^2 \Bigl(2\, m_0^2 \
\frac{\partial f_1}{\partial q^2}+ 2\, m_0  \frac{\partial f_1}{\partial m}+f_1\Bigr)+2\, c_2\, c_1 \Bigl(2\, m_0^2 \
\frac{\partial f_0}{\partial q^2}+ 2\, m_0\,\frac{\partial f_0}{\partial m}+f_0\Bigr) \nonumber \\ 
&\qquad +2\,\left(c_3\, f_0+c_2\, f_1+c_1\, f_2\right) m_0  \biggr) \biggr|_{q^2=m_0^2} \nonumber\\
= \biggl. & \alpha_s^3 \biggl(c_1^3\, \Bigl(\frac{4}{3}\, m_0^3\, \frac{\partial^2 f_0}{\partial (q^2)^2}+2\, m_0^2\, \frac{\partial^2 f_0}{\partial q^2 \,\partial m}+m_0 \
\frac{\partial^2 f_0}{\partial m^2} %+\frac{\partial f_0}{\partial m}
\Bigr) + c_1^2 \Bigl(2\, m_0^2 \,\frac{\partial f_1}{\partial q^2} + 2\, m_0 \, \frac{\partial f_1}{\partial m}\Bigr)
%+4\, c_2\, c_1\, m_0\, \frac{\partial f_0}{\partial m}
\biggr) \biggr|_{q^2=m_0^2}\,,
\end{align}
where $m_0$ is a $b$-quark mass different from $\mbos$ and $c_i$ are the perturbative coefficients in the mass-conversion relation between $\mbos$ and $m_0$, as defined in eq.~\eqref{eq:transformation}. 
The second reduced form follows from taking into account the vanishing of, at least, the following pieces for the $q^2$-spectrum of $b \rightarrow u \ell \bar{\nu}_\ell$: 
\begin{align}
f_0(m_0^2, m_0) = f_1(m_0^2, m_0) = f_2(m_0^2, m_0) = \Big. \frac{\partial f_0}{\partial q^2} \Big|_{m = m_0,q^2=m_0^2} = \Big. \frac{\partial f_0}{\partial m} \Big|_{m = m_0,q^2=m_0^2} = 0\,.
\end{align}
Under this condition, we see that the boundary-effect terms for the $q^2$-spectrum start to become non-vanishing for $b \rightarrow u \ell \bar{\nu}_\ell$ but only from the third-order in $\als$.

If these boundary-effect terms are not vanishing, where shall we attribute them in terms of the distribution $\tilde{f}(\alpha_s\,,\, m_0\,;\, x)$ defined in \eqref{eq:integralexp}? 
Apparently, they shall be viewed as the additional contribution associated with the boundary bin of the histogrammed integration domain covering $x=m_0$ and formally beyond. 
In the limit of an infinitesimal binning, the non-zero boundary-effect terms will manifest themselves in the form of Dirac-$\delta$ type modification to the distribution $\tilde{f}(\alpha_s\,,\, m_0\,;\, x)$ around $x=m_0$.

We note that the existence of these boundary derivatives in \eqref{eq:boundaryterm} are implicitly \textit{assumed} for the perturbative expressions at least up to the order in question. 
Otherwise, the above perturbative re-expansion, and hence $\tilde{f}(\alpha_s\,,\, m_0\,;\, x)$,  simply does not exist around the expanded boundaries. 
We note in addition that, for the $q^2$-spectrum in the semi-leptonic decays of heavy-light mesons, in particular $\BXulv$, the fixed-order perturbative results alone are generally insufficient for making reliable physical predictions in the vicinity of the edge of the maximum invariant mass.
The technical point revealed above concerns merely the high-order perturbative QCD corrections, the focus of the present work.
~\\

The point revealed above has a profound impact on our following presentation of the re-expanded perturbative QCD corrections to the $q^2$-spectrum defined with quark masses other than the pole mass: 
given the non-vanishing boundary-effect terms, it does not appear to us to be sensible --- or we do not know how --- to plot a truly differential $q^2$-spectrum in $\MSbar$-renormalized mass scheme with the scale-variation bands including the expanded $\overline{m}(\mu)$-dependent boundary of the partonic phase space, not even in principle. %(let alone the additional technical issue of scale dependence of $\MSmass(\mu)$ in the resulting boundary of the expanded partonic phase space).
On the other hand, the integrated (inclusive) moments of the distribution typically have much simpler quark-mass dependence, and consequently allow much more straightforward transformation of mass schemes. 
Given our discussions above, we thus limit ourselves to presenting the perturbatively reformulated $q^2$-spectrum with $\mu$-scale variations in the \textit{histogram} form with finite bin width --- the inclusive moments may be viewed as the special case with just a single binning ---
%unless explicitly stated otherwise for illustration purpose, and 
in a few representative $\mu$-independent threshold-mass schemes.
~\\

However, let us emphasize that in the remaining text of this section we are mainly concerned with demonstrating the results of the high-order perturbative corrections using different quark-mass schemes. 
A comprehensive calculation and phenomenological discussion, e.g.~incorporated the relevant non-perturbative power corrections dedicated to different differential observables, is beyond the scope envisaged for the present work.

\subsubsection{$q^2$-spectrum for $b \rightarrow u\,\ell \bar{\nu}_{\ell}$}
\label{sec:lpmass_b2ulv}

\begin{figure}[htbp]
\begin{center}
\includegraphics[scale=0.6]{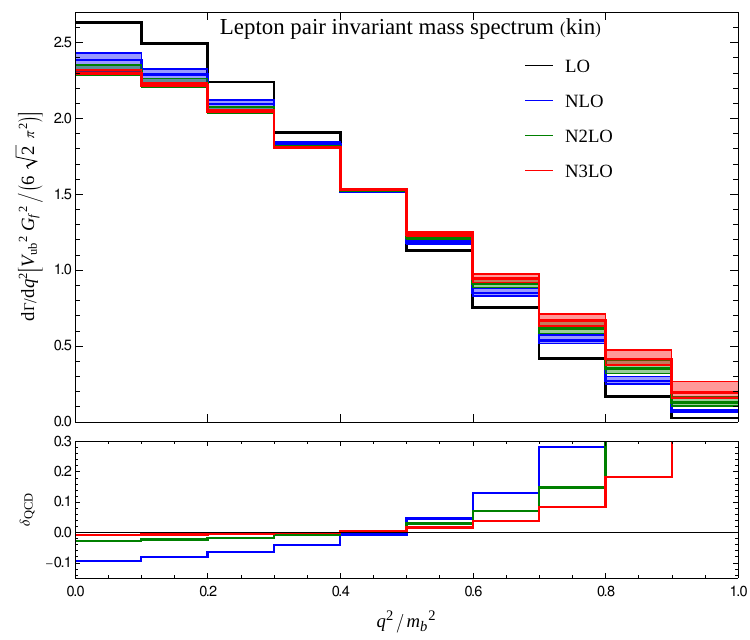}
\includegraphics[scale=0.6]{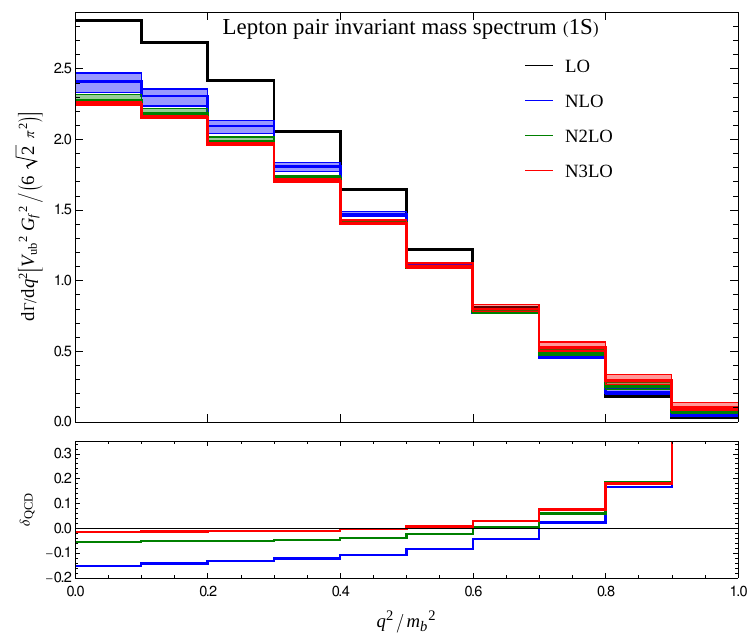}
\caption{Left: the perturbative results for the histogrammed $q^2$-spectrum determined in the kinetic mass scheme at different orders in $\alpha_s$ with the conventional scale variation $\mu \in [\mbkin/2, \mbkin]$. The horizontal x-axis is the rescaled $q^2$ normalized by $\mbkin^2$, hence the absolute range of $q^2$ in this plot is different from that in fig.~\ref{fig:b2ulv_dGamdq2_OS}. %A constant factor is pulled out as the unit for the values shown on the y-axis. 
Right: the perturbative results for the $q^2$-spectrum determined in the 1S mass scheme plotted in the same way as the left plot. 
(Note that in this plot the $q^2$ in x-axis is normalized by $\mbOs^2$, and hence the bin width and range in $q^2$ is different from the left plot.)
}
\label{fig:b2ulv_dGamdq2_Kin_1S}
\end{center}
\end{figure}

Having addressed the tricky point involved in perturbatively re-expanding the quark-mass  dependence of a differential distribution with consistent integrated results, we are now ready to present our perturbative results for the $q^2$-spectrum in $\bulv$ reformulated using alternative threshold mass schemes. 
Given the analysis on $\GamBulv$ in section~\ref{sec:Bdecay}, we first consider the kinetic mass~\cite{Bigi:1994ga,Bigi:1996si} $\mbkin$ for $b$-quark. 
Using the input $\mbkin[1] \equiv \mbkin(\mu_c=1\,\mathrm{GeV}) = 4.554 \pm 0.018$~GeV from the HFLAV collaboration~\cite{HFLAV:2019otj}, our novel numerical results for the histogrammed $q^2$-spectrum determined in the kinetic mass scheme are shown in the left plot in fig.~\ref{fig:b2ulv_dGamdq2_Kin_1S}.
This result for $q^2$-spectrum histogrammed in terms of 10 bins for $q^2 \in [0, \mbkin^2]$\footnote{In the fully-fledged application to $\BXulv$, the maximum value of $q^2$ shall be determined by the physical masses of the hadrons, independent of the intermediate quark masses in principle. %affecting the actual bin width of the last bin
} was derived from the on-shell result shown in fig.~\ref{fig:b2ulv_dGamdq2_OS} using the matching relation between the pole mass and kinetic mass up to three-loop order~\cite{Fael:2020iea}, following the treatment explained in the previous subsection~\ref{sec:lpmass_reformula}.
In particular, the boundary-effect term derived in~\eqref{eq:boundaryterm} is not vanishing for $q^2$-spectrum in $\bulv$ but only from $\mathcal{O}(\als^3)$. 
Its contribution has been included in the right-most bin at $q^2/ \mbkin^2 = 1$ and affects the value in this bin by $3 \sim 5\%$ (depending on the choice of scales).
We note that even without this boundary-effect term, the value of the $q^2$-spectrum in the vicinity of the  boundary $q^2/ \mbkin^2 = 1$ starts to deviate from 0 from $\mathcal{O}(\als^2)$, due to the involvement of the high-order mass-derivatives in the first part of the r.h.s.~of \eqref{eq:integralpowexp}.
More comments on this will be given at the end of this subsection.

As noted earlier, changing to alternative quark-mass schemes for the \textit{inclusive} moments of the $q^2$-spectrum is as straightforward as for the full inclusive decay width, which is nothing but actually the $0$-th moment. 
This is because the $m_b$ dependence of these inclusive moments takes a very simple monomial form with the power determined by their mass dimensions (as all other particles are taken massless). 
We have checked explicitly the equivalence between the two sets of the inclusive moments of the $q^2$-spectrum derived respectively, up to weight $q^6$, by (1) integrating directly the $q^2$-spectrum derived in the kinetic mass scheme, whose histogrammed form is given in fig.~\ref{fig:b2ulv_dGamdq2_Kin_1S}; 
and by (2) transforming the inclusive moments in the on-shell scheme derived from fig.~\ref{fig:b2ulv_dGamdq2_OS} following the simple procedure as done for the inclusive decay width $\GamBulv$. 
We emphasize that the equivalence is observed but only after incorporating the above boundary-effect term.

Since the kinetic mass $\mbkin$ is independent of $\als$-renormalization scale $\mu$, the leading-order result in fig.~\ref{fig:b2ulv_dGamdq2_Kin_1S} is just one line without any conventional scale-variation bound.\footnote{In this preliminary study of the $q^2$-spectrum without incorporating any non-perturbative power contributions involving $1/m_b$-suppressed high-dimensional operators, we refrain from investigating further the dependence of these perturbative results on the Wilson-cutoff dependence of the $\mbkin$.}
Compared to the on-shell results for $q^2$-spectrum in fig.~\ref{fig:b2ulv_dGamdq2_OS}, the most notable difference shown in the new results in fig.~\ref{fig:b2ulv_dGamdq2_Kin_1S} is that the size of the QCD corrections is reduced considerably, at each order in $\alpha_s$.
Accordingly, the size of the conventional scale-variation bands, obtained by varying $\mu$ from $\mbkin/2$ to $\mbkin$, is reduced compared to those in the on-shell scheme; 
and now their sizes are also reduced as higher-order corrections are incorporated, as usually expected.  
Moreover, they start to cluster together and overlap especially in the low-$q^2$ region. 
On the other hand, the fixed-order perturbative corrections are generally unstable and unreliable in the vicinity of the maximum $q^2$ end-point, which can also be seen from the quick rising of the $\delta_{\mathrm{QCD}}$ at each order shown in the lower panel.

A closer look at the relative QCD corrections, shown by $\delta_{\mathrm{QCD}}$ in the lower panel, reveals another interesting feature: 
the sign of $\delta_{\mathrm{QCD}}$ flips as one moves from the lower-$q^2$ to the high-$q^2$ region at each order in $\alpha_s$, crossing zero at about the middle region of the $q^2$-domain. 
This particular crossing pattern thus offers us an illuminating perspective to appreciate the previously puzzling feature exhibited in the QCD corrections to the inclusive $\GamBulv$ determined in the kinetic mass scheme, in fig.~\ref{fig:TotWidthMuD_MSvsKin}: the conventional scale variations are very small and their sizes do not decrease much as one goes to higher orders (especially from N2LO to N3LO). 
There is thus no contradiction between the above feature in $\GamBulv$, i.e.~the inclusive 0-th moment of the $q^2$-spectrum, and the regular or usual behavior of $\delta_{\mathrm{QCD}}$ shown in fig.~\ref{fig:b2ulv_dGamdq2_Kin_1S}.

On the other hand, when restricted to the large-$q^2$ region, say $q^2 > (m_b-m_c)^2$~\cite{Bauer:2000xf}, which is about $10~\text{GeV}^2$ around the middle of the histogram plot, 
the $\mathcal{O}(\als^3)$ corrections are quite sizable, clearly shown in Fig.~\ref{fig:b2ulv_dGamdq2_Kin_1S}. 
To be specific, at the Born-level, the sum of the first 5 bins in the $q^2$-histogram, covering $q^2 \in (0, 10.36)~$GeV$^2$, is $4.33$ times of those in the remaining 5 bins.
The QCD correction factor up to $\Oals{3}$ for the first 5 bins in the low-$q^2$ region in total reads 
$1 - 0.06371 -0.01586 -0.004016 = 0.916$ with the conventional scale uncertainty $[-0.2 \%, 0.5 \%]$ which is very good. 
However, it reads for the remaining contribution in the large-$q^2$ region,
$1 \,+\, 0.1647 \,+\, 0.1264 \,+\, 0.09519 = 1.386$ with the conventional scale uncertainty $[-4 \%, 6\%]$ in relative.
Therefore, the effect of the $\Oals{3}$ perturbative QCD corrections to the $q^2$-spectrum in the large-$q^2$ region is significantly larger than in the full inclusive moments (including the total decay width~\eqref{eq:GB2Xulv}), even in the kinetic mass scheme, and thus deserves extra care. 
The large-$q^2$ region had played an important role in the inclusive determination of $|V_{ub}|$ at Belle~\cite{Belle:2003vfz,HFLAV:2016hnz,Belle-II:2018jsg,Belle-II:2022cgf}, especially using the approaches of refs.~\cite{Bauer:2000xf,Bauer:2001rc}.
The perturbative QCD uncertainties constitute an important source of errors in this approach\cite{Belle:2003vfz}, albeit not listed as the largest in the past analysis~\cite{HFLAV:2016hnz}.  
Given our discussion above, it would be interesting to investigate the potential impact of our findings in clarifying the puzzling large differences in the inclusive determination of $|V_{ub}|$ extracted using this approach as compared to using the others~\cite{HFLAV:2016hnz,Belle-II:2018jsg}, and similarly the persistent tension between the inclusive and exclusive determination of $|V_{ub}|$~\cite{HFLAV:2019otj,HFLAV:2022esi} in future fully fledged analysis, e.g.~, whether the previous theoretical uncertainties were underestimated to some extent. 
~\\

In addition to the kinetic mass scheme, we considered, for comparison, another threshold quark mass definition that is frequently employed in the study of heavy-light mesons, the 1S mass~\cite{Hoang:1998ng,Hoang:1999zc}. 
In the right plot in fig.~\ref{fig:b2ulv_dGamdq2_Kin_1S}, our novel numerical results for the histogrammed $q^2$-spectrum determined in the 1S mass are shown, using the input $\mbOs = 4.67$~GeV from RunDec and the mass conversion relations provided in ref.~\cite{Hoang:1998ng,Hoang:1999zc}.
The results are plotted in the same way as in the left plot for the results in the kinetic mass scheme, but note the different range in $q^2$ between these two plots. 
Just as the left plot, it exhibits a similar, much improved convergence behavior of the QCD corrections to the $q^2$-spectrum after reformulation from the original on-shell mass scheme. 
It is worthy to notice that compared to the results in the kinetic mass scheme, the effect of the perturbative QCD corrections in the large-$q^2$ region is slightly smaller in the 1S mass scheme. 
To be specific, at the Born-level, the sum of the first 5 bins in the $q^2$-histogram, covering $q^2 \in (0, 10.90)~$GeV$^2$, is again $4.33$ times of those in the remaining 5 bins.
The QCD correction factor up to $\Oals{3}$ or third-order according to the expansion prescription of the 1S mass scheme~\cite{Hoang:1998ng,Hoang:1999zc} for the first 5 bins in the low-$q^2$ region in total reads 
$1-0.1339 -0.0431 -0.00964 = 0.813$ with the conventional scale uncertainty $[-0.2 \%, 0.5 \%]$.
However, it reads for the remaining contribution in the large-$q^2$ region,
$1 -0.0297 + 0.0245 + 0.0539 = 1.05$ with the conventional scale uncertainty $[-4 \%, 6\%]$. 
Our results show that the effect of the third-order perturbative QCD corrections to the $q^2$-spectrum in the large-$q^2$ region is much more significant than in the full inclusive moments in both the kinetic and 1S mass schemes, and thus deserves extra care for studies with cuts to this region.   
~\\

Before leaving this subsection, we would like to add a quick comment on the endpoint behavior of the perturbatively re-expanded $q^2$-spectrum with a non-pole mass defintion for the heavy quark. 
To better illustrate this point, we show the plots in the $\sigma$-mass definition~\cite{Chen:2025iul,Chen:2025zfa}, employed when rewriting $\GamTsl$ for the $t$-quark decay in the previous subsection~\ref{sec:Tdecay}. 
Although $\sigma$-mass is a short-distance mass for quarks free of the leading IR-renormalon issue, being independent of renormalization scale and scheme in addition, it is, however, not really a quasi-on-shell or threshold mass in the sense that its numerical value is not close to the pole mass as the kinetic mass or 1S mass. 
Its numerical value for $b$-quark, $\mbsm = 3.96$~GeV, is, instead, numerically closer to $\mbms(\mbms)$ albeit not exactly the same.
This makes it less advantageous to be employed in the framework of HQET for decay processes of $b$-quark and $c$-quark at relatively low energies, where the specifically designed threshold masses, e.g.~kinetic mass or 1S mass, are more favored.
Despite this, owing to its \textit{independence} on the $\als$-renormalization scale $\mu$, it might be interesting and convenient to plot the perturbative results for $q^2$-spectrum in this mass scheme, especially in view of the absence of similar plots using the $\MSbar$-mass scheme. 
Another technical reason for this is that given the relatively small value $\mbsm = 3.96$~GeV, compared to the pole mass $\mbos$ and quasi-on-shell $\mbkin$ and $\mbOs$, the change of the end-point behavior of the perturbatively re-expanded $q^2$-spectrum becomes more pronounced and visible --- the point of this discussion --- as will be clear in a moment.

\begin{figure}[htbp]
\begin{center}
\includegraphics[scale=0.6]{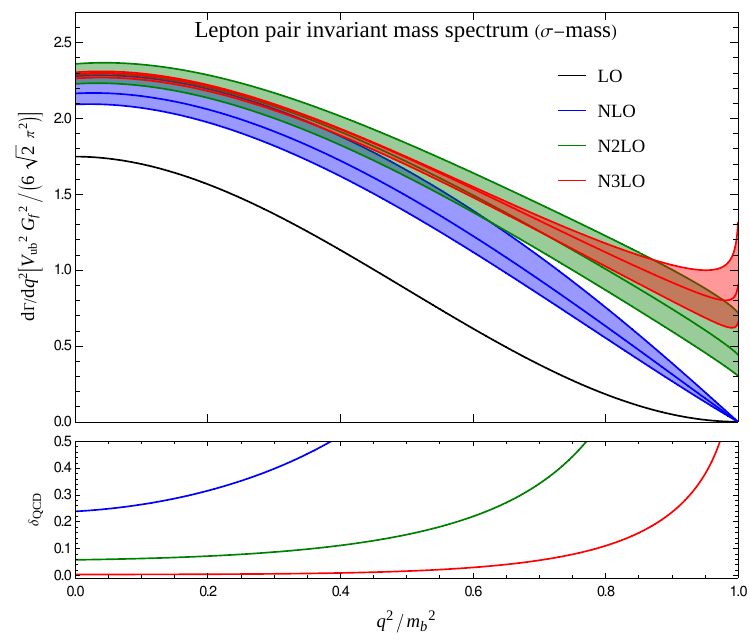}
\includegraphics[scale=0.6]{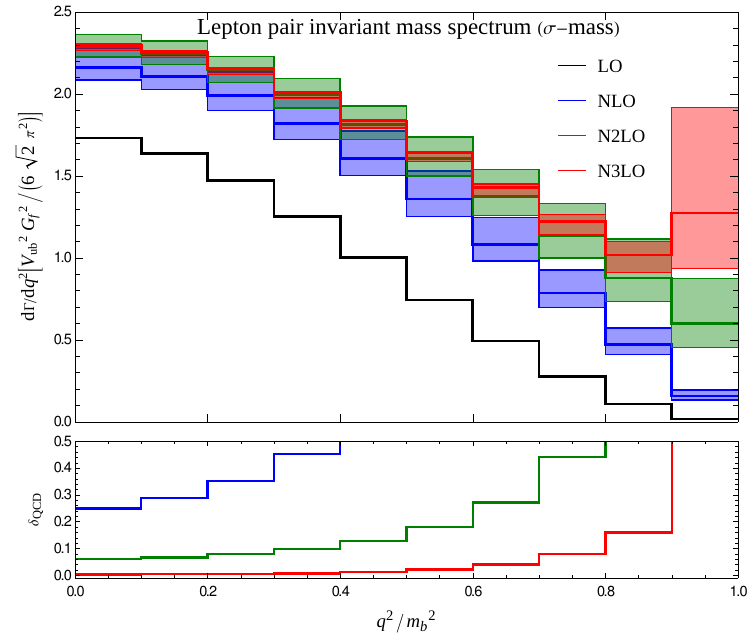}
\caption{Left: the perturbative results for the $q^2$-spectrum determined in the $\sigma$-mass scheme at different orders in $\alpha_s$ with the conventional scale variation $\mu \in [\mbsm/2, \mbsm]$ \textit{excluding} the boundary of the expanded phase-space. 
The horizontal x-axis is the rescaled $q^2$ normalized by $\mbsm^2$. %A constant factor is pulled out as the unit for the values shown on the y-axis.
Right: the histogrammed counterpart of the left plot but with the boundary-effect terms in~\eqref{eq:boundarytermRefine} included in the right-most bin.  
}
\label{fig:b2ulv_dGamdq2_sigM}
\end{center}
\end{figure}

Our numerical results for the $q^2$-spectrum determined in the $\sigma$-mass scheme at different orders in $\alpha_s$ are shown in fig.~\ref{fig:b2ulv_dGamdq2_sigM} with the conventional scale variations. 
In the left plot, we produce a continuous version for the $q^2$-spectrum \textit{without}  incorporating the boundary-effect term defined in~\eqref{eq:boundaryterm}, which is non-zero only from $\mathcal{O}(\alpha_s^3)$ and can not be done anyway.  
We now see clearly that the value of the continuous $q^2$-spectrum starts to deviate from 0 from $\mathcal{O}(\als^2)$ in the vicinity of the end-point $q^2/ \mbsm^2 = 1$, due to the involvement of the high-order mass-derivatives in the first part of the r.h.s.~of \eqref{eq:integralpowexp}.
In the right plot, the histogrammed counterpart of the left plot is provided, but with the boundary-effect term included in the right-most bin;   
due to the relatively small value $\mbsm = 3.96$~GeV, the value in this bin is increased by about 34\%.

It is not surprising to observe, generally speaking, the increased perturbative QCD correction factors and associated scale-variation bands in fig.~\ref{fig:b2ulv_dGamdq2_sigM} compared to the corresponding results derived in the kinetic or 1S mass scheme shown in fig.~\ref{fig:b2ulv_dGamdq2_Kin_1S}, especially at lower perturbative orders.
On the other hand, there is a clear reduction in the scale uncertainty as one moves to higher orders. 
Both the size of $\mathcal{O}(\alpha_s^3)$ correction and the residual scale uncertainty at this perturbative order are actually  reduced to levels comparable to those shown in the aforementioned results in the low-$q^2$ region. 
Furthermore, it is more clearly shown in this plot that the fixed-order perturbative results alone are generally unstable in the vicinity of the end-point of the $q^2$-domain.

\subsubsection{$q^2$-spectrum for $\cdlv$}
\label{sec:lpmass_c2dlv}

The semi-leptonic decay of a $c$-quark, e.g.~$\cdlv$, in a 4-flavor QCD with only $c$-quark massive involves virtually the same set of Feynman diagrams as the $\bulv$, except for having one quark flavor less. 
Therefore, our results for the heavy-to-light structure functions $W_i$ can also be applied to $\cdlv$.
On the other hand, the relatively large value of $\als$ at the typical energy scale of $c$-quark decay,  given by its mass $\sim 1.27$~GeV, $\als(\mcms) \approx 0.38$ which is about twice of $\als(\mbms)$ --- again almost twice of $\als(\mtms)$ --- naturally leads to concerns regarding the rate of the convergence of the high-order perturbative corrections. 
The knowledge of higher-order perturbative QCD corrections is clearly longed-for in a recent comprehensive study~\cite{King:2021xqp} of D-mesons lifetimes, of their ratios and of the inclusive semileptonic decay rates.
Irrespective of whether the knowledge of higher-order perturbative QCD corrections to $\cdlv$ is useful to actually resolve the issues reported in ref.~\cite{King:2021xqp}, the explicit results may be of help to clarify, in the first place, the question of convergence and hence applicability of the perturbative QCD calculation for this process. 
On the other hand, in view of the low-energy scale of the semi-leptonic decay of D-mesons and the relatively larger number of non-perturbative HQET parameters (due to the slow convergence rate of the $1/m_c$-power expansion), it is quite reassuring to know that there have been recently very promising progress on the first-principles calculation of this decay process from lattice QCD~\cite{Hashimoto:2017wqo,Gambino:2020crt,Gambino:2022dvu,DeSantis:2025qbb,DeSantis:2025yfm,Kellermann:2025pzt}, producing results that agree with experiments within errors.

Below we will conduct some tentative applications of our perturbative results of the heavy-to-light structure functions $W_i$ to the $q^2$-spectrum in $\cdlv$ up to $\mathcal{O}(\als^3)$ for the first time, and to the inclusive moments of the lepton-energy spectrum in the next subsection.
Let us note that the moments of $q^2$-spectrum in inclusive semileptonic decays of D-mesons are reparametrization invariant and quite sensitive to non-perturbative contributions parameterized in HQET~\cite{Gambino:2005tp}. 
Their precise measurements at the BES III experiment~\cite{Asner:2008nq} might help to disentangle the various non-perturbative effects.

Let us begin by documenting the input parameters used in the numerical results. 
For the masses of $c$-quark, we take $\mcms(\mcms) = 1.27 \pm 0.02$~GeV, $\mcOs = 1.55$~GeV~\cite{ParticleDataGroup:2024cfk}, 
and $\mckin = 1.40(7)$~GeV at the Wilsonian cutoff scale $\mu_c=0.5$~GeV\cite{Gambino:2010jz}.
To determine the values of $\als$ around the $c$-quark mass, we take the input $\als(\mu=\mcms) = 0.38$ as the initial point in a four-loop running with 4 flavors.
First of all, the perturative QCD corrections to the total decay width of $\cdlv$ determined with $\MSbar$ mass are large and show a very slow convergence.
To get an idea, the relative sizes of the perturative QCD corrections at $\mu=\mcms$ normalized by the tree-level result read $\{0.52, 0.44, 0.42 \}$.

\begin{figure}[htbp]
\begin{center}
\includegraphics[scale=0.6]{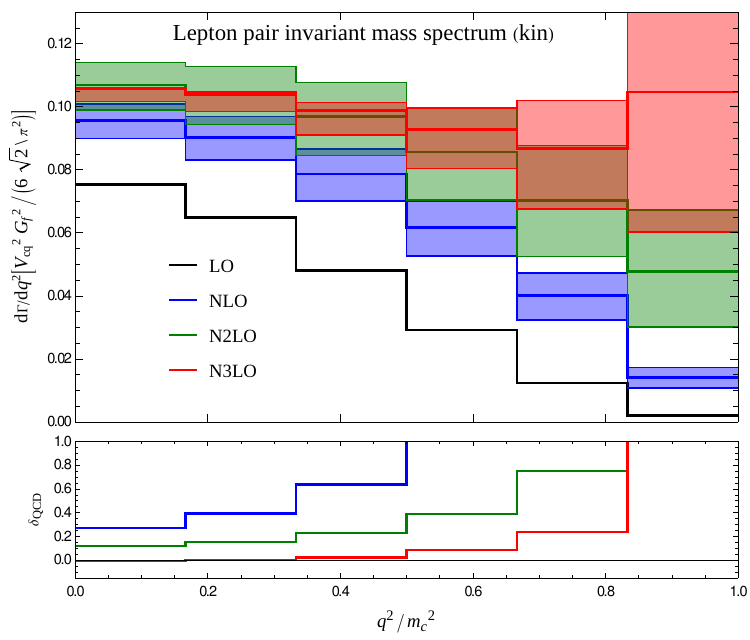}
\includegraphics[scale=0.6]{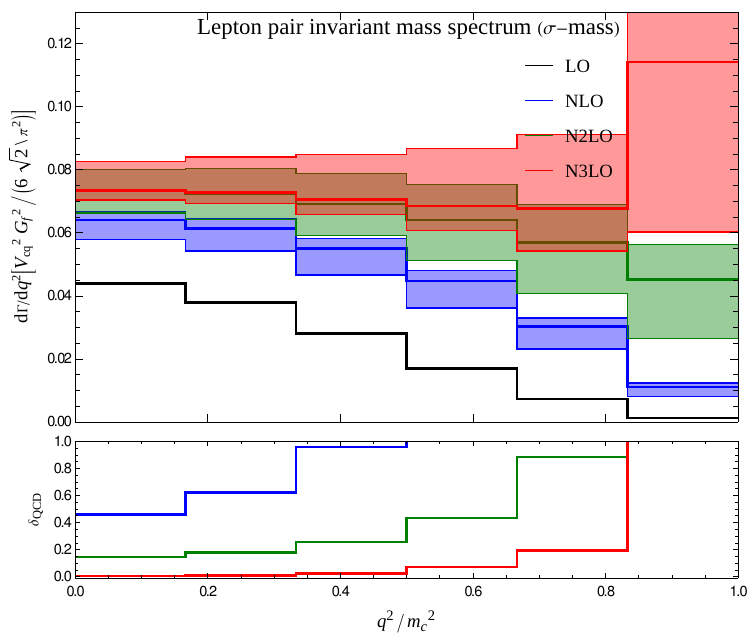}
\caption{Left: the perturbative results for the histogrammed $q^2$-spectrum determined in the kinetic mass scheme at different orders in $\alpha_s$ with the scale variation $\mu \in [1~\text{GeV}, 2\,\mckin]$. The horizontal x-axis is the rescaled $q^2$ normalized by $\mckin$. 
Right: the perturbative results for the $q^2$-spectrum determined in the $\sigma$-mass scheme plotted in the same way as the left plot. (Note that in this plot the $q^2$ in x-axis is normalized by $\mbsm^2$, and hence the bin width and range in $q^2$ are different from the left plot. 
)}
\label{fig:c2dlv_dGamdq2_Kin_sigM}
\end{center}
\end{figure}

Following the same procedure discussed in the previous subsection on $\bulv$, we determined the perturbative results for the $q^2$-spectrum in the kinetic mass scheme for $\cdlv$, which are shown in the left plot of fig.~\ref{fig:c2dlv_dGamdq2_Kin_sigM} in the histogram form with 5 bins in $q^2 \in [0, \mckin^2]$. 
We see that unlike the previous application to $\bulv$, the perturbative QCD corrections do not seem to converge even in the kinetic mass scheme, except perhaps in the low-$q^2$ region, e.g.~with $q^2 < \mckin^2/2$.  
Note that the lower scale of the scale-variation bands in this plot was simply set at $1$~GeV, rather than $\mckin/2$ which, in our opinion, is too low for sensible applications of the perturbative QCD.
For comparison, we have also included a similar plot for the perturbative corrections to the $q^2$-spectrum determined using the $\sigma$-mass scheme with $\mcsm = 1.17$~GeV.
(Note that the bin widths in $q^2$ are actually different between these two plots, due to the different values of $\mckin$ and $\mcsm$.)

\begin{figure}[htbp]
\begin{center}
\includegraphics[scale=0.75]{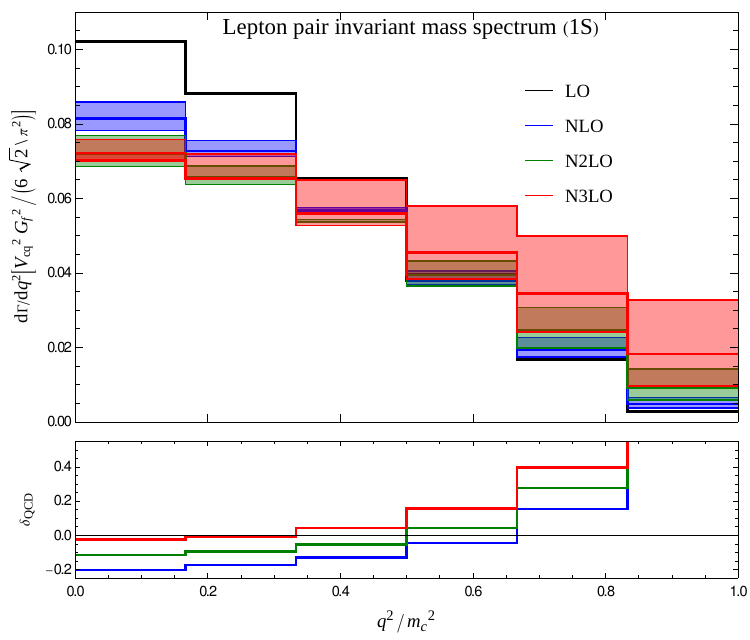}
\caption{The perturbative results for the histogrammed $q^2$-spectrum determined in the 1S mass scheme at different orders in $\alpha_s$ with the scale variation $\mu \in [1~\text{GeV}, 2\,\mcOs]$. The horizontal x-axis is the rescaled $q^2$ normalized by $\mcOs$.
}
\label{fig:c2dlv_dGamdq2_1S}
\end{center}
\end{figure}

It is interesting to observe that the perturbative QCD corrections determined in the 1S mass scheme up to $\mathcal{O}(\alpha_s^3)$, shown in fig.~\ref{fig:c2dlv_dGamdq2_1S}, seem to behave better, at least, in the low-$q^2$ region with $q^2 < \mcOs^2/2$, for reasons that are not clear to us at the moment.
However, it may be convenient to recall at this moment that the expansion prescription of the 1S mass scheme~\cite{Hoang:1998ng} is not a pure power-series expansion in $\als$, unlike the cases of the other short-distance and/or threshold mass schemes considered before.
Our preliminary investigation of the perturbative QCD corrections to the $q^2$-spectrum of $\cdlv$  indicates that with a suitable short-distance or threshold mass, notably the 1S mass, the perturbative QCD corrections to the observables defined for the semi-leptonic inclusive decays of D-meson may still be useful and applied, at least to $\mathcal{O}(\als^3)$, under a kinematic cut $q^2 < \mcOs^2/2$.

\subsection{Moments of the electron-energy spectrum in $\cqlv$}
\label{sec:eem}

The lepton-energy spectrum, or its moments, in the inclusive semi-leptonic decay of D-mesons are important precision observables in experiments that are sensitive to the $c$-sector of the CKM matrix elements and several non-perturbative hadronic parameters.
Both the semi-leptonic decay width and electron-energy moments (EEM) have been measured precisely for $D_s \rightarrow X\,\ell \bar{\nu}_{\ell}$ and the current experimental precision now reaches the level of a few ($2\sim 3$) \%~\cite{BESIII:2021duu,DeSantis:2025qbb}.
Very recently, the first-principles calculations of these inclusive moments from lattice QCD have been completed~\cite{DeSantis:2025qbb,DeSantis:2025yfm}, where the central values of the theoretical predictions agree with the experimental measurements~\cite{CLEO:2009uah,BESIII:2021duu} within errors,  although the total errors of the theoretical results are still larger than the experimental ones~\cite{BESIII:2021duu}.  
From the perturbation theory side, so far only the $\mathcal{O}(\als)$~\cite{Aquila:2005hq} and $\mathcal{O}(\als^2\, \beta_0)$~\cite{Czarnecki:1994pu,Lange:2005yw,Gambino:2007rp} corrections are readily available for these observables, and are employed in the previous analysis, e.g.~in refs.\cite{Gambino:2010jz,Gambino:2022dvu}.\footnote{We kindly note that the Monte-Carlo set-up~\cite{Brucherseifer:2013cu} can be employed in principle to compute the complete $\mathcal{O}(\als^2)$ corrections to these inclusive moments, at least in the on-shell scheme.}
In refs.~\cite{Shao:2025vhe,Shao:2025qwp} the first determination of the CKM matrix elements $V_{cs}$ and $V_{cd}$, together with HQET non-perturbative parameters, from a global fit to data from inclusive D-meson decays, had been performed, where our perturbative QCD corrections up to $\mathcal{O}(\als^2)$ to the inclusive EEMs played an important role in improving the precision of the fit.

Below we will report our new perturbative results for the electron-energy spectrum in $\cqlv$ up to $\mathcal{O}(\als^3)$.
Due to the well-known endpoint behavior of the lepton-energy spectrum~\cite{Bigi:1992su,Bigi:1993fe,Bigi:1993ex,Neubert:1993ch,Neubert:1993um,Manohar:2000dt,Lange:2005yw} in heavy-to-light semi-leptonic decays (with massless final-state quarks), such as $\cqlv$, 
rather than attempting to plot the full energy spectrum either in the continuous or histogram form, we will be computing only their inclusive EEMs up to weight 2.
(Limited to just QCD corrections, the dependence of the triple-differential distribution on lepton energy $E_l$ in eq.~\ref{eq:3Dparametric} is a second-order polynomial.)
Another practical advantage of this is that the subsequent transition to short-distance and threshold quark-mass schemes for the inclusive moments is as straightforward as for the full  inclusive decay width, owing to the simple monomial dependence on the $c$-quark mass. 
To be definite, we define the following perturbative (un-normalized) EEMs for $\cqlv$: 
\begin{eqnarray}\label{eq:EEMs_c2qlv}
\langle E_e^{N} \rangle
&=&\int^{E_e^{\mathrm{max}}}_{0} 
E_e^{N} \, \frac{\mathrm{d} \GamBulv }{\mathrm{d}\, E_e} \, \mathrm{d}\, E_e\,
\equiv \frac{G_F^2\, |V_{cq}|^2}{192 \, \pi^3} \, \langle E_e^{N} \rangle\big|_{unn.}
\end{eqnarray}
where $E_e^{\mathrm{max}}$ equals the pole mass $\mcos$ in the on-shell scheme (with massless quarks in the final state). 
And in the second equality, the overall factor $\frac{G_F^2\, |V_{cq}|^2}{192 \, \pi^3}$ collecting the CKM matrix elements is pulled out for the convenience in presenting the numerical values below (No electroweak K-factor is included here).
In light of the discussions in the preceding subsections, we will present the numerical results for $\cqlv$ only in the 1S mass scheme.
For the input parameters used, we take $\mcOs = 1.55$~GeV~\cite{ParticleDataGroup:2024cfk}, and $\als(\mu=\mcOs) = 0.338$, which is determined via four-loop $\alpha_s$-running with the initial value $\als(\mu=\mcms) = 0.38$~\cite{ParticleDataGroup:2024cfk}.  
We are now ready to present the numerical results for the perturbative EEMs at $\mu = \mcOs$ in \eqref{eq:eem_result_central}: 
\begin{eqnarray}\label{eq:eem_result_central}
\langle E_e^{0} \rangle \big|_{unn.} &=& 8.9466 - 1.1871 ~\epsilon - 0.4440 ~\epsilon^2 - 0.0609 ~\epsilon^3 \,+\, \mathcal{O}(\epsilon^4) \nonumber\\
\langle E_e^{1} \rangle \big|_{unn.} &=& 4.1602 - 0.4994  ~\epsilon - 0.1307 ~\epsilon^2 + 0.0861 ~\epsilon^3 \,+\, \mathcal{O}(\epsilon^4) \nonumber\\
\langle E_e^{2} \rangle \big|_{unn.} &=& 2.1494 - 0.2311 ~\epsilon - 0.02498 ~\epsilon^2 + 0.1107 ~\epsilon^3 \,+\, \mathcal{O}(\epsilon^4) \,,
\end{eqnarray}
where the $\epsilon$-tag is a formal expansion parameter introduced in the 1S-mass expansion prescription to track the order of perturbation theory~\cite{Hoang:1998ng,Hoang:1999zc}. 
More specially, the $\Oals{N}$ term in the perturbative series of the decay width is assigned $\epsilon^N$ while the $\Oals{N}$ term (modular logarithmic dependence) in the pole-to-1S mass conversion relation is regarded as $\epsilon^{N-1}$;
after re-expansion and truncation to a given order in $\epsilon$, this parameter will be set 1 in the end. 
The 0-th EEM in \eqref{eq:eem_result_central} is precisely the inclusive decay width for $\cqlv$, expanded to third order in $\epsilon$. 
Although the structure function $W_5$ is not involved in the 0-th EEM, it contributes to other EEMs. 
The formal 1S-expansion parameter $\epsilon$ shall be set to 1 to resume the actual results:
\begin{eqnarray}\label{eq:eem_results}
\langle E_e^{0} \rangle \big|_{unn.} = 7.26(37) \,, \quad 
\langle E_e^{1} \rangle \big|_{unn.} = 3.62(37) \,, \quad 
\langle E_e^{2} \rangle \big|_{unn.} = 2.00(30) \,,
\end{eqnarray}
where the numbers in brackets give the variation of the results as the $\alpha_s$-renormalization scale $\mu$ is varied from $\mcms = 1.27$~GeV to $2\,\mcOs$.

The signs of the perturbative QCD corrections in \eqref{eq:eem_result_central} start to change from the $\epsilon^2$ to the next $\epsilon^3$ term for EEM apart from the 0-th order one, i.e.~the inclusive decay width.
If this indicates a zero point in the $\epsilon$-coefficient --- viewed as a continuous function of the 1S-expansion order around 2 or 3 in the surrounding region --- then it may provide a unified perspective for understanding the smallness of the $\epsilon$-coefficients shown in \eqref{eq:eem_result_central}; 
however, the apparently good convergence behavior observed in these formal series may thus not necessarily persist at higher orders in $\epsilon$. 
On the other hand, as the QCD perturbation series is known to be generally asymptotic, the central practical point lies simply in determining the optimal truncation order to extract an accurate and reliable estimate from the available information for a given observable.

\section{Conclusion}
\label{sec:conc}

In this work we present the first complete calculation of all five heavy-to-light structure functions for semi-leptonic weak decays of heavy quarks up to $\mathcal{O}(\alpha_s^3)$ in QCD, establishing a new perturbative precision frontier for these processes. 
This is achieved by employing a specifically-designed hybrid strategy to solve bivariate master integrals that combines an efficient linear interpolation based on stratified Gauss-Kronrod points in the $q^2$-region with deeply-expanded series in the other variable(s), further armed with reduced numerical $\varepsilon$-dependence. 
The perturbative results for the triple-differential semi-leptonic decay rates of heavy-quarks thus finally become available at the unprecedented accuracy of $\mathcal{O}(\alpha_s^3)$.

For demonstration of example applications,  we have focused on a few selected precision observables in the semi-leptonic heavy-to-light decays of $t$-,~$b$-and $c$-quark, evaluated using different quark-mass schemes, which are crucial for extracting the relevant CKM elements and/or non-perturbative OPE parameters from the experiments.
In the $t$-decay case, we derived the reliable theoretical prediction $\Gamt = 1.321(3)(2) \times|V_{tb}|^2 + 0.027\,(m_t - 172.69)\,\text{GeV}$, whose precision meets the requirements of future colliders. 
In application to $\BXulv$, by combining our new $\mathcal{O}(\alpha_s^3)$ perturbative corrections with all percent-level non-perturbative corrections in the kinetic-mass scheme, a state-of-the-art prediction is obtained:
$\Gamma(\BXulv) =  \frac{|V_{ub}|^2}{|3.82\times 10^{-3}|^2}\,\big( 6.53 \,\pm 0.12 \, \pm 0.13\, \pm 0.03\, \big) \times 10^{-16}\,\text{GeV}\,$. 
This result is of significant value, both for the ambitious goal of achieving percent-level precision in the inclusive determination of $|V_{ub}|$ at Belle II and for addressing the long-standing tension between its inclusive and exclusive measurements.
For $\cdlv$, we have had the first look at the perturbative results for the inclusive moments of the electron-energy spectrum up to $\mathcal{O}(\als^3)$, and interestingly, the size of perturbative QCD corrections seem to remain under control in the 1S-mass scheme, at least, up to this order. 
In view of the recent simultaneous fit of the CKM matrix elements $|V_{cs}|$ and $|V_{cd}|$ together with non-perturbative parameters that utilize these inclusive moments measured at BES III, we look forward to seeing the impact of our higher-order results.

In addition to the above high-order perturbative results for selected precision observables, we have reported several notable observations regarding the convergence behavior of the perturbative QCD corrections to the inclusive decay widths and the differential $q^2$-spectrum. 
We have paid particular attention to the impact of changing to different short-distance and threshold quark-mass definitions.
The information revealed through these investigation are both interesting and valuable, 
especially in view of the fact that the characteristic energy scales of heavy-meson decays are not very high, and thus a good convergence behavior is crucial not only for the precision of the perturbative results but also for a reliable estimate of related systematic errors. 
In particular, we observed that in both the kinetic and 1S mass schemes employed in extracting $|V_{ub}|$, the perturbative QCD corrections to the $q^2$-spectrum in $\bulv$ flip sign as moving from the lower-$q^2$ to the high-$q^2$ region at each order in $\alpha_s$, crossing zero around the middle region of the $q^2$-domain. 
This particular crossing pattern offers an illuminating perspective to appreciate a puzzling feature exhibited in the QCD corrections to the inclusive $\GamBulv$ determined in the kinetic mass scheme: the conventional scale variations are very small and moreover their sizes do not decrease much as moving to higher orders (especially from N2LO to N3LO). 
On the other hand, the effect of the $\Oals{3}$ perturbative QCD corrections to the $q^2$-spectrum in the large-$q^2$ region, say $q^2 > (m_B-m_D)^2$, is significantly larger than in the full inclusive moments (including the inclusive decay width), both in the kinetic and 1S mass scheme. 
Given the important role of the large-$q^2$ region in the inclusive determination of $|V_{ub}|$ at Belle and Belle II, it would be interesting to investigate the potential contribution of our findings in clarifying the puzzling large differences in the inclusive determination of $|V_{ub}|$ using different approaches (focusing on different kinematic regions), and similarly the persistent tension between the inclusive and exclusive determination of $|V_{ub}|$ in future fully fledged analysis.

One more interesting point worthy of being noted in our study is a technical subtlety involved in the consistent perturbative reformulation of the $q^2$-spectrum at the full differential level when changing from the pole-mass scheme to other mass schemes (e.g., $\overline{\text{MS}}$, kinetic and 1S mass). 
We identified and calculated the boundary-effect terms that arise from the perturbative re-expansion of the phase-space integration limit itself, explicitly for the differential $q^2$-spectrum; 
they are non-vanishing for $\bulv$ but only from $\mathcal{O}(\alpha_s^3)$ onward, whose incorporation are essential to preserve the integrity of integrated moments of the perturbatively re-expanded differential spectrum.
As a consequence of this, we conclude that as far as the truncated perturbative QCD corrections are concerned, a fully differential $q^2$-spectrum reformulated in a short-distance or threshold mass scheme following the usual prescription does not exist for $\bulv$ beyond $\Oals{2}$ around the expanded boundaries of the partonic phase space; 
and one has go to the histograms which is demanded even within the fixed-order perturbation theory alone.

In this work, we have primarily focused on demonstrating the results of high-order perturbative corrections and applications to a few selected observables using different quark-mass schemes. 
A comprehensive analysis and detailed discussion of their full impact on phenomenology --- after incorporating the relevant known non-perturbative corrections for experimentally measured observables --- are reserved for future studies. 
In addition, we note that the mass effect of final-state charged lepton, e.g.~in $b \rightarrow u \ell \bar{\nu}_{\ell}$ with $l = \mu,\, \tau$, can be easily accounted for by adapting the final-state phase space, and thus the $W_i$ presented in this work can also be used to study these semi-leptonic decay channels.   
Moreover, the hybrid strategy employed to compute $W_i$ that combines an efficient linear interpolation (using a suitable function basis that needs not be purely polynomial) based on stratified Gauss-Kronrod points in one dimension and the deeply-expanded series in the other degree(s) of freedom --- further armed with reduced numerical $\varepsilon$-dependence --- can also be applied to $\BXclv$ to take into account the mass effect of the final-state quarks.  
In summary, our calculation provides the state-of-the-art N3LO perturbative corrections to all hadronic structure functions in heavy-to-light semi-leptonic weak decays, representing a major advancement in the perturbative precision of theoretical predictions for these processes. 
With the improved predictions for various double- and single-differential decay rates of $\BXulv$ and $\Dqlv$ becoming available soon, our results are poised to significantly improve the precision of inclusive determinations of CKM matrix elements, such as $|V_{ub}|$, $|V_{cs}|$, and $|V_{cd}|$ and of non-perturbative parameters (including shape-functions) in ongoing experiments at BES III, Belle II, and LHCb.

\section*{Acknowledgements}
The work of L.C. was supported by the Natural Science Foundation of China under contract No.~12205171, No.~12235008 and No.~12321005, and Department of Science and Technology of Shandong province No.~tsqn202312052 and~2024HWYQ-005.
X.C. was supported by the Swiss National Science Foundation (SNF) under contract 200020\_219367 and the UZH Postdoc Grant, grant no. [FK-25-104].
X.G. was supported by the United States Department of Energy, Contract DE-AC02-76SF00515. 
Y.Q.M. was supported in part by the National Natural Science Foundation of China under contract No.~12325503. 
~\\

\textbf{Note Added: }
When finalizing this paper, we noticed that the other group~\cite{Broggio:2026edk} had just completed a semi-analytic extraction of the hadronic structure functions $W_1, W_3, W_5$ (relevant for $\bulv$ with massless leptons) at $\Oals{2}$. In the region with low $q^2 < 0.5 m_Q^2$ and relatively large $p\cdot q /m_Q$, where the numerical error in Ref.~\cite{Broggio:2026edk} is small, we find very good agreement for the pure $\Oals{2}$ corrections to $W_1, W_3, W_5$, with the relative difference below 1\%.

\bibliographystyle{utphysMa}
\bibliography{HeavQuarkDecay}

\end{document}